\pdfoutput=1
\documentclass[11pt,twoside,a4paper,cmspaper,final,collab]{cms-tdr}
\def\svnVersion{fc7e758-D}\def\svnDate{2021/08/05}\def\cmsCernNoTag{CERN-EP-2020-234}\def\cmsCernDate{\today}\def\cmsMessage{"Published in Physics Letters B as \href{http://dx.doi.org/10.1016/j.physletb.2021.136565}{\doi{10.1016/j.physletb.2021.136565}.}"}
\begin{document}\cmsNoteHeader{TOP-20-003}

\newcommand{\ttcc}{{\ttbar\PQc\PAQc}}
\newcommand{\ttcL}{{\ttbar\PQc\text{L}}}
\newcommand{\ttbb}{{\ttbar\PQb\PAQb}}
\newcommand{\ttbL}{{\ttbar\PQb\text{L}}}
\newcommand{\ttLF}{{\ttbar\text{LL}}}
\newcommand{\ttjj}{{\ttbar\text{jj}}}

\hyphenation{TensorFlow}

\ifthenelse{\boolean{cms@external}}{\providecommand{\cmsLeft}{upper\xspace}}{\providecommand{\cmsLeft}{left\xspace}} 
\ifthenelse{\boolean{cms@external}}{\providecommand{\cmsRight}{lower\xspace}}{\providecommand{\cmsRight}{right\xspace}} 

\providecommand{\cmsTable}[1]{\resizebox{0.5\textwidth}{!}{#1}}

\renewcommand*\descriptionlabel[1]{\hspace\labelsep\normalfont #1}

\ifthenelse{\boolean{cms@external}}{\newcommand{\appen}{~}}{\newcommand{\appen}{Appendix~}}

\cmsNoteHeader{TOP-20-003} 
\title{First measurement of the cross section for top quark pair production with additional charm jets using dileptonic final states in \texorpdfstring{$\Pp\Pp$}{pp} collisions at \texorpdfstring{$\sqrt{s} = 13\TeV$}{sqrt(s) = 13 TeV}}

\author*[inst1]{The CMS Collaboration}

\date{\today}

\abstract{
   The first measurement of the inclusive cross section for top quark pairs ($\ttbar$) produced in association with two additional charm jets is presented. The analysis uses the dileptonic final states of $\ttbar$ events produced in proton-proton collisions at a centre-of-mass energy of 13\TeV. The data correspond to an integrated luminosity of 41.5\fbinv, recorded by the CMS experiment at the LHC. A new charm jet identification algorithm provides input to a neural network that is trained to distinguish among $\ttbar$ events with two additional charm ($\ttcc$), bottom ($\ttbb$), and light-flavour or gluon ($\ttLF$) jets. By means of a template fitting procedure, the inclusive $\ttcc$, $\ttbb$, and $\ttLF$ cross sections are simultaneously measured, together with their ratios to the inclusive $\ttbar$ + two jets cross section. This provides measurements of the $\ttcc$ and $\ttbb$ cross sections of $10.1\pm 1.2 \stat\pm 1.4\syst\unit{pb}$ and $4.54\pm 0.35\stat\pm 0.56\syst\unit{pb}$, respectively, in the full phase space. The results are compared and found to be consistent with predictions from two different matrix element generators with next-to-leading order accuracy in quantum chromodynamics, interfaced with a parton shower simulation.
}

\hypersetup{
pdfauthor={CMS Collaboration},
pdftitle={First measurement of the cross section for top quark pair production with additional charm jets using dileptonic final states in pp collisions at sqrt(s) = 13 TeV},
pdfsubject={CMS},
pdfkeywords={CMS, top quark, dilepton, charm quark, charm-tagging, heavy-flavour}}

\maketitle

\section{Introduction}\label{sec:Introduction}

The modelling of top quark pair ($\ttbar$) production in association with jets from the hadronization of bottom ({\PQb}) or charm ({\PQc}) quarks, referred to as {\PQb} jets and {\PQc} jets, respectively, in proton-proton ($\Pp\Pp$) collisions at the CERN LHC is challenging. Calculations of the production cross section for top quark pairs with additional pairs of {\PQb} jets ($\ttbb$) are available at next-to-leading order (NLO) in quantum chromodynamics (QCD)~\cite{Bredenstein:2008zb,Bredenstein:2009aj,Cascioli:2013era,Jezo:2018yaf,Buccioni:2019plc}, but suffer from large uncertainties due to the choice of factorization ($\mu_{\text{F}}$) and renormalization ($\mu_{\text{R}}$) scales. The uncertainties arise from the different energy (or mass) scales in $\ttbb$ production, which range from the large scales associated with the top quark mass ($m_{\PQt}$), to the relatively small scales associated with additional jets resulting mostly from gluon splitting into $\PQb\PAQb$ pairs. To better understand the $\ttbb$ process, the ATLAS and CMS experiments have conducted several measurements in $\Pp\Pp$ collisions at centre-of-mass energies of 7, 8, and $13\TeV$ \cite{Aad:2013tua,CMS:2014yxa,Aad:2015yja,Khachatryan:2015mva,Sirunyan:2017snr,Aaboud:2018eki,Sirunyan:2019jud,Sirunyan:2020kga}. 

The production of a $\ttbar$ pair with an additional pair of {\PQc} jets ($\ttcc$) has received far less attention, both theoretically and experimentally. Whereas the experimental signature of a {\PQb} jet looks quite different from that of a light-flavour (LF) or gluon jet, the differences are much less pronounced for c jets. This explains the challenge of simultaneously separating $\ttcc$ and $\ttbb$ events from a large background of $\ttbar$ events with additional LF or gluon jets ($\ttLF$). With the development of a charm jet identification algorithm ("{\PQc} tagger") \cite{Sirunyan:2017ezt}, these signatures can now be more efficiently disentangled. We present the first measurement of the inclusive $\ttcc$ cross section and its ratio to the inclusive $\ttbar$ + two jets ($\ttjj$) cross section. A fully consistent treatment of the different additional jet flavours is ensured using a technique that simultaneously extracts the $\ttcc$, $\ttbb$, and $\ttLF$ cross sections. The measurement is performed in the dileptonic decay channel of the $\ttbar$ events using a data sample of $\Pp\Pp$ collisions at a centre-of-mass energy of $13\TeV$, collected with the CMS detector in 2017, corresponding to an integrated luminosity of 41.5\fbinv~\cite{CMS-PAS-LUM-17-004}. This data set benefits from the upgrade of the pixel tracking detector~\cite{CMS-TDR-011}, which was installed in winter 2016--2017 and which has been shown to significantly improve the performance of heavy-flavour (HF) jet identification~\cite{CMS-DP-2018-033}.

Although the additional {\PQb} jets in $\ttbb$ events are predominantly produced via gluon splitting into $\PQb\PAQb$ pairs, they can also originate from the decay of a Higgs boson (\PH). Previous measurements of Higgs boson production in association with a top quark pair, where the Higgs boson decays into a pair of {\PQb} quarks ($\ttbar\PH$, $\PH\to \PQb\PAQb$)~\cite{Sirunyan:2018hoz,Sirunyan:2018mvw,Aaboud:2018urx,Aaboud:2017rss}, suffer from a nonresonant background of gluon-induced $\ttbb$ events, and to a lesser extent also from $\ttcc$ events due to the misidentification of {\PQc} jets as {\PQb} jets. The techniques described here provide a basis for a simultaneous measurement of the $\ttbb$ and $\ttcc$ processes from data that can be adopted in future $\ttbar\PH$ analyses to significantly reduce the uncertainties related to these backgrounds.

\section{The CMS detector}\label{sec:detector}

The central feature of the CMS apparatus is a superconducting solenoid of 6\unit{m} internal diameter, providing a magnetic field of 3.8\unit{T}. Within the solenoid volume are a silicon pixel and strip tracker, a lead tungstate crystal electromagnetic calorimeter (ECAL), and a brass and scintillator hadron calorimeter (HCAL), each composed of a barrel and two endcap sections. The silicon tracker measures charged particles within the pseudorapidity range $\abs{\eta} < 2.5$. During the LHC running period when the data used for this analysis were recorded, the silicon tracker consisted of 1856 silicon pixel and 15\,148 silicon strip detector modules. For nonisolated particles of $1 < \pt < 10\GeV$ and $\abs{\eta} < 2.5$, the track resolutions are typically 1.5\% in \pt and 20--75\mum in the transverse impact parameter~\cite{CMS-DP-2020-032}. Forward calorimeters extend the $\eta$ coverage provided by the barrel and endcap detectors. Muons are detected in gas-ionization chambers embedded in the steel flux-return yoke outside the solenoid. A more detailed description of the CMS detector, together with a definition of the coordinate system used and the relevant kinematic variables, can be found in Ref.~\cite{Chatrchyan:2008zzk}.

Events of interest are selected using a two-tiered trigger system~\cite{Khachatryan:2016bia}. The first level, composed of custom hardware processors, uses information from the calorimeters and muon detectors to select events at a rate of around 100\unit{kHz} within a fixed time interval of about 4\mus. The second level, known as the high-level trigger, consists of a farm of processors running a version of the full event reconstruction software optimized for fast processing, and reduces the event rate to around 1\unit{kHz} before data storage.

\section{Event simulation}\label{sec:Simulation}

Samples of signal and background events are simulated using Monte Carlo (MC) event generators based on a fixed-order perturbative QCD calculation with up to four noncollinear high-$\pt$ partons, supplemented with parton showering (PS) and multiparton interactions.
The matrix element (ME) generation of $\ttbar$ events is performed with \POWHEG (v2)~\cite{Nason:2004rx,Frixione:2007vw,Alioli:2010xd, Campbell:2014kua,Frixione:2007nw} at NLO in QCD using the five-flavour scheme, followed by a simulation of the PS using $\PYTHIA8.230$~\cite{Sjostrand:2014zea} (referred to as $\PYTHIA8$ in the following), using the \textsc{CP5} underlying event tune~\cite{Sirunyan:2019dfx} and the NNPDF3.1~\cite{Ball:2017nwa} parton distribution functions (PDFs). The first additional hard radiation from the $\ttbar$ system is included in the NLO ME calculation, whereas higher additional jet multiplicities result from radiation simulated in the PS. A value of $m_{\PQt} = 172.5\GeV$ is used in the event generation and the values of $\mu_{\text{R}}$ and $\mu_{\text{F}}$ are set to a dynamic scale given by $\sqrt{\smash[b]{m^2_{\cPqt} + p^2_{\mathrm{T,\cPqt}}}}$, where $p_{\mathrm{T},\cPqt}$ denotes the transverse momentum of the top quark in the $\ttbar$ rest frame. The $\ttbar$ cross section is scaled to a theoretical prediction at next-to-next-to-leading order in QCD including resummation of next-to-next-to-leading logarithmic soft-gluon terms, which yields $\sigma_{\ttbar} = 832^{\, +39.9}_{\, -45.8}$\unit{pb}~\cite{Cacciari:2011hy}. The results are also compared to those from $\ttbar$ production simulated with the \MGvATNLO (v2.4.2)~\cite{Alwall:2014hca} ME generator at NLO accuracy using the five-flavour scheme, with FxFx jet matching~\cite{Frederix:2012ps} and including up to two jets in addition to the $\ttbar$ system in the NLO ME calculation. No dedicated simulations have been used to model separately $\ttcc$ or $\ttbb$ events, ensuring a consistent treatment between $\ttcc$, $\ttbb$, and $\ttLF$ events in the inclusive $\ttbar$ samples mentioned above.

The background processes for this analysis consist mainly of Drell--Yan (DY) and single top quark events, with additional minor contributions from {\PW}$+\,$jets, diboson, triboson, $\ttbar\PZ$, $\ttbar\PW$, and $\ttbar\PH$ events (collectively referred to as rare backgrounds). For all these samples, the PS is simulated using $\PYTHIA8$. The ME generation of the DY, {\PW}$+\,$jets, triboson, $\ttbar\PZ$, and $\ttbar\PW$ events is handled using \MGvATNLO at leading order in QCD, with MLM jet matching~\cite{Alwall:2007fs}.
The $\ttbar\PH$ events are generated at NLO using \POWHEG. Single top quark production in the $t$ and $s$ channels are simulated at NLO in the four-flavour scheme using \POWHEG and \MGvATNLO, respectively, while the $\PQt\PW$ channel is simulated at NLO with \POWHEG in the five-flavour scheme. 
The diboson samples are simulated at leading order in QCD using $\PYTHIA8$ for both the ME calculations and PS description. 

The interactions between particles and the material in the CMS detector are simulated using \GEANTfour (v10.02)~\cite{Agostinelli:2002hh}. The effect of additional $\Pp\Pp$ interactions in the same or nearby beam crossings as the hard-scattering process (pileup) is modelled by adding simulated minimum bias collisions generated in $\PYTHIA8$.

\section{Signal definition}\label{sec:signal}

\begin{figure}[th!]
\centering
\includegraphics[width=0.49\textwidth]{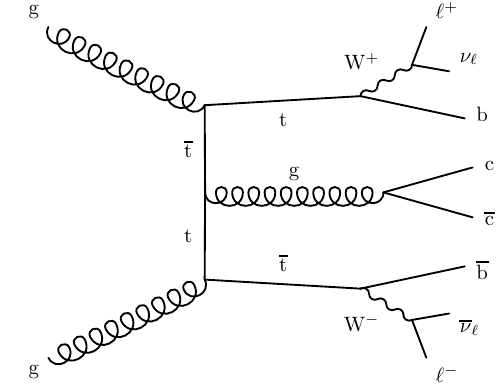}
\caption{\label{fig:Feynmandileptonttcc}
Example of a Feynman diagram at the lowest order in QCD, describing the dileptonic decay channel of a top quark pair with an additional $\PQc\PAQc$ pair produced via gluon splitting.}
\end{figure}

A typical Feynman diagram describing the dileptonic $\ttcc$ process is shown in Fig.~\ref{fig:Feynmandileptonttcc}. To provide an unambiguous interpretation of the results, a generator-level definition of the event categories is needed, based on the flavours of the additional jets. The HF jets are identified at the generator level using the procedure of ghost matching~\cite{Cacciari:2007fd}, where in addition to the reconstructed final-state particles, the generated {\PQb} and {\PQc} hadrons are clustered into the jets. However, the modulus of the hadron four-momentum is set to a small number to prevent these generated hadrons from affecting the reconstructed jet momentum and to ensure that only their directional information is retained. Jets that contain both {\PQb} and {\PQc} hadrons are labelled as {\PQb} jets, since the {\PQc} hadrons most likely originate from a decay of a {\PQb} hadron. The results of this measurement are reported in terms of the fiducial and full phase spaces.

\textit{Fiducial phase space}: In the definition of the fiducial phase space, all of the final-state particles (except for the neutrinos) resulting from the decay chain: $\Pp \Pp \to \ttbar\text{jj} \to \ell^{+} \nu_{\ell} \PQb  \ell^{-}\overline{\nu}_{\ell} \PAQb \text{jj}$ are required to be within the region of the detector in which these objects can be properly reconstructed. The fiducial phase space is therefore defined by the presence of two oppositely charged electrons, muons, or $\tau$ leptons ($\ell$) at the generator level with $\pt > 25\GeV$ and $\abs{\eta} < 2.4$. 
Each lepton is required to originate from a decay of a \PW boson, which in turn results from a top quark decay. Particle-level jets are defined by clustering generated final-state particles with a mean lifetime greater than 30~ps, excluding neutrinos, using the anti-\kt algorithm~\cite{Cacciari:2008gp, Cacciari:2011ma} with a distance parameter of 0.4. We demand two particle-level {\PQb} jets from the top quark decays with $\pt > 20\GeV$ and $\abs{\eta} < 2.4$. Besides these two {\PQb} jets, at least two additional particle-level jets must be present, with the same kinematic requirements imposed. Jets that lie within $\Delta\text{R}=\sqrt{\smash[b]{(\Delta\eta)^2+(\Delta\phi)^2}} < 0.4$ of either one of the leptons from the \PW boson decays, where $\phi$ denotes the azimuthal angle, are excluded. The above requirements define the inclusive $\ttbar\text{jj}$ signal, which is then subdivided based on the flavour content of the additional particle-level jets (not from the top quark decays) into the following categories:
\begin{description}
\item[$\ttbb$:] At least two additional {\PQb} jets are present, each containing at least one {\PQb} hadron.
\item[$\ttbL$:] Only one additional {\PQb} jet is present, containing at least one {\PQb} hadron. In addition, at least one additional LF or {\PQc} jet is present. This category results from $\ttbb$ events in which one of the two additional {\PQb} jets is outside the acceptance, or two {\PQb} jets are merged into one.
\item[$\ttcc$:] No additional {\PQb} jets are present, but at least two additional {\PQc} jets are found, each containing at least one {\PQc} hadron.
\item[$\ttcL$:] No additional {\PQb} jets are present, and only one additional {\PQc} jet is found, containing at least one {\PQc} hadron. In addition, at least one additional LF jet is present. This category results from $\ttcc$ events in which one of the two additional {\PQc} jets is outside the acceptance, or two {\PQc} jets are merged into one.
\item[$\ttLF$:] No additional {\PQb} or {\PQc} jets are present, but at least two additional LF jets are within the acceptance.
\end{description}

All other $\ttbar$ events that do not fit in any of the above categories, because they do not fulfill the acceptance requirements described in the definition of the fiducial phase space, are labelled as ``$\ttbar$+other". These could for example be events with semileptonic or fully hadronic $\ttbar$ decays that pass the event selection criteria outlined in Section~\ref{sec:selection}, or dileptonic $\ttbar$ events for which the leptons or the particle-level jets as described above are not within the fiducial detector volume. These contributions are estimated from the same simulations as those used for the signal events.

\textit{Full phase space}: The definition of the full phase space comprises dileptonic, semileptonic, and fully hadronic $\ttbar$ decays that contain in addition at least two particle-level jets with $\pt > 20\GeV$ and $\abs{\eta} < 2.4$. These jets must not originate from the decays of the top quarks or \PW bosons. There are no requirements imposed on the generator-level leptons or on the particle-level jets that result from the top quark and \PW boson decays. The measurement in the fiducial phase space is extrapolated to the full phase space by applying an acceptance factor, estimated from simulations, to each signal category based on the additional jet flavours.

The definition of the fiducial phase space is much closer to the reconstructed phase space in which the measurement is performed, and therefore expected to suffers less from theoretical uncertainties that affect the extrapolation from the fiducial to the full phase space. However, the full phase space definition is better suited for comparison with theoretical calculations.

\section{Object reconstruction}\label{sec:ObjectReconstruction}

The global event reconstruction is based on the particle-flow algorithm~\cite{CMS-PRF-14-001}, which aims to reconstruct and identify each individual particle in an event by combining information from the various elements of the CMS detector. The energy of photons is obtained from the ECAL measurement. The energy of electrons is determined from a combination of the electron momentum at the primary interaction vertex as determined by the tracker, the energy of the corresponding ECAL cluster, and the energy sum of all bremsstrahlung photons spatially compatible with originating from the electron track. The energy of muons is obtained from the curvature of the corresponding track. The energy of charged hadrons is determined from a combination of their momentum measured in the tracker and the matching ECAL and HCAL energy deposits, corrected for zero-suppression effects and for the response function of the calorimeters to hadronic showers. Finally, the energy of neutral hadrons is obtained from the corresponding corrected ECAL and HCAL energies. 

For each event, hadronic jets are clustered from these reconstructed particles using the infrared and collinear safe anti-\kt algorithm with a distance parameter of 0.4. Jet momentum is determined as the vectorial sum of all particle momenta in the jet, and is found from simulations to be, on average, within 5 to 10\% of the true momentum over the whole $\pt$ spectrum and detector acceptance. Pileup interactions can contribute additional tracks and calorimetric energy depositions, increasing the apparent jet momentum. To mitigate this effect, tracks identified as originating from pileup vertices are discarded and an offset correction is applied to correct for remaining contributions. Jet energy corrections are derived from simulation studies so that the average measured response of jets becomes identical to that of particle-level jets. In situ measurements of the momentum balance in dijet, photon+jets, {\PZ}$\,+\,$jets, and multijet events are used to determine any residual differences between the jet energy scale in data and simulations~\cite{Khachatryan:2016kdb}, and appropriate corrections are made. Additional selection criteria are applied to each jet to remove jets potentially dominated by instrumental effects or reconstruction failures. 

The missing transverse momentum vector \ptvecmiss is computed as the negative of the vector sum of the \ptvec of all the particle-flow candidates in an event, and its magnitude is denoted as \ptmiss~\cite{Sirunyan:2019kia}. The \ptvecmiss is modified to account for corrections to the energy scale of the reconstructed jets in the event.

The candidate vertex with the largest value of summed physics-object $\pt^2$ is taken to be the primary $\Pp\Pp$ interaction vertex. The physics objects are the jets and the associated \ptmiss.

Jets originating from the hadronization of {\PQb} and {\PQc} quarks are identified using the deep combined secondary vertex (DeepCSV) algorithm~\cite{Sirunyan:2017ezt}, which uses information on the decay vertices of long-lived mesons and the impact parameters of the charged particle tracks as input to a deep neural network (NN) classifier. For the identification of {\PQb} jets, a medium working point is chosen, corresponding to a ${\approx}70\%$ efficiency for correctly selecting {\PQb} jets and a misidentification probability for LF ({\PQc}) jets of ${\approx}1 \, (12)\%$, derived from simulated $\ttbar$ events. The same algorithm also provides discriminators to distinguish {\PQc} jets from LF and {\PQb} jets, which are collectively referred to as {\PQc} tagging discriminators. The information from the distributions of these observables plays a key role in this analysis, and the calibration of the {\PQc} tagger is further discussed in Section~\ref{sec:ctagCalib}.

\section{Event selection}\label{sec:selection}

An event selection has been employed to select a subset of events that consists almost exclusively (more than 95\% as evaluated from simulations) of dileptonic $\ttbar$ events with at least two additional jets. Exactly two reconstructed, oppositely charged leptons (either electrons or muons) are required to be present. This procedure also selects $\PGt$ leptons that decay into an electron or a muon. The electrons and muons are required to have $\pt > 25\GeV$ and $\abs{\eta} < 2.4$, and should be isolated from other objects in the event.
At least four jets are required in the event, all of which must be spatially separated from the isolated leptons by imposing $\Delta \text{R}(\ell,\text{jet}) > 0.5$. Only jets with $\pt > 30\GeV$ and $\abs{\eta} < 2.4$ are considered. An assignment of the jets to the expected partons is made to identify the {\PQb} jets from top quark decays and jets originating from additional radiation (described in detail in Section~\ref{sec:matching}). The two jets assigned to the {\PQb} quarks from the top quark decays are required to be {\PQb} tagged. Neutrinos from leptonically decaying \PW bosons are not detected, but instead contribute to the \ptmiss, which is required to exceed $30\GeV$ in events with two electrons ($\Pe\Pe$) or two muons ($\PGm\PGm$), in order to reduce contributions from DY events. In events with one electron and one muon ($\Pe\PGm$), no requirement is imposed on \ptmiss. In order to further reduce the contribution from DY production in $\Pe\Pe$ and $\PGm\PGm$ events, the invariant mass of the two leptons ($m_{\ell\ell}$) is required to be outside of the \PZ boson mass window, $m_{\ell\ell} \not\in [ m_{\PZ} - 15,  m_{\PZ} + 15 ]\GeV$, with $m_{\PZ} = 91.2\GeV$~\cite{PDG2020}. For all events, it is required that $m_{\ell\ell}>12\GeV$ in order to minimize contributions from low-mass resonances.

\section{Matching jets to partons}\label{sec:matching}

The distinction among the $\ttcc$, $\ttbb$, and $\ttLF$ categories relies on the correct identification of the additional jets not coming from the decay of the top quarks. Assuming a 100\% branching fraction for the decay $\PQt\to\PQb\PW$, and focusing on the dileptonic decay channel, two {\PQb} jets are expected from the top quark decays and at least two additional jets are required through the other event selection criteria. In practice, not all {\PQb} jets from top quark decays will be reconstructed within the acceptance of the detector and additional jets will also not necessarily pass the reconstruction and selection criteria. In this section, a matching procedure is described to achieve the most accurate correspondence between the final-state jets and the expected partons. This is done by considering all possible permutations of four jets in the collection of jets passing the selection criteria described in Section~\ref{sec:selection} and training a NN to identify the correct jet-parton assignment. 

Whether a given permutation corresponds to a correct jet-parton assignment can be inferred from different quantities, such as the jet kinematic variables, {\PQb} and {\PQc} tagging discriminators, and angular separations and invariant masses between pairs of jets (and between jets and leptons). The NN processes, for each event, all possible jet-parton assignments and is trained on the aforementioned observables to give the highest possible score to the correct permutation. For the $\ttLF$, $\ttcc$, and $\ttcL$ categories, the {\PQb} and {\PQc} tagging discriminators dominate the decision made by the NN. For the $\ttbb$ category all four partons have the same flavour, so the correct assignment of the four {\PQb} jets relies mainly on the angular separations and invariant masses between jets or leptons. The best assignment for the $\ttbL$ category benefits from a combination of all these observables. 

\begin{figure}[ht!]
\centering
\includegraphics[width=0.49\textwidth]{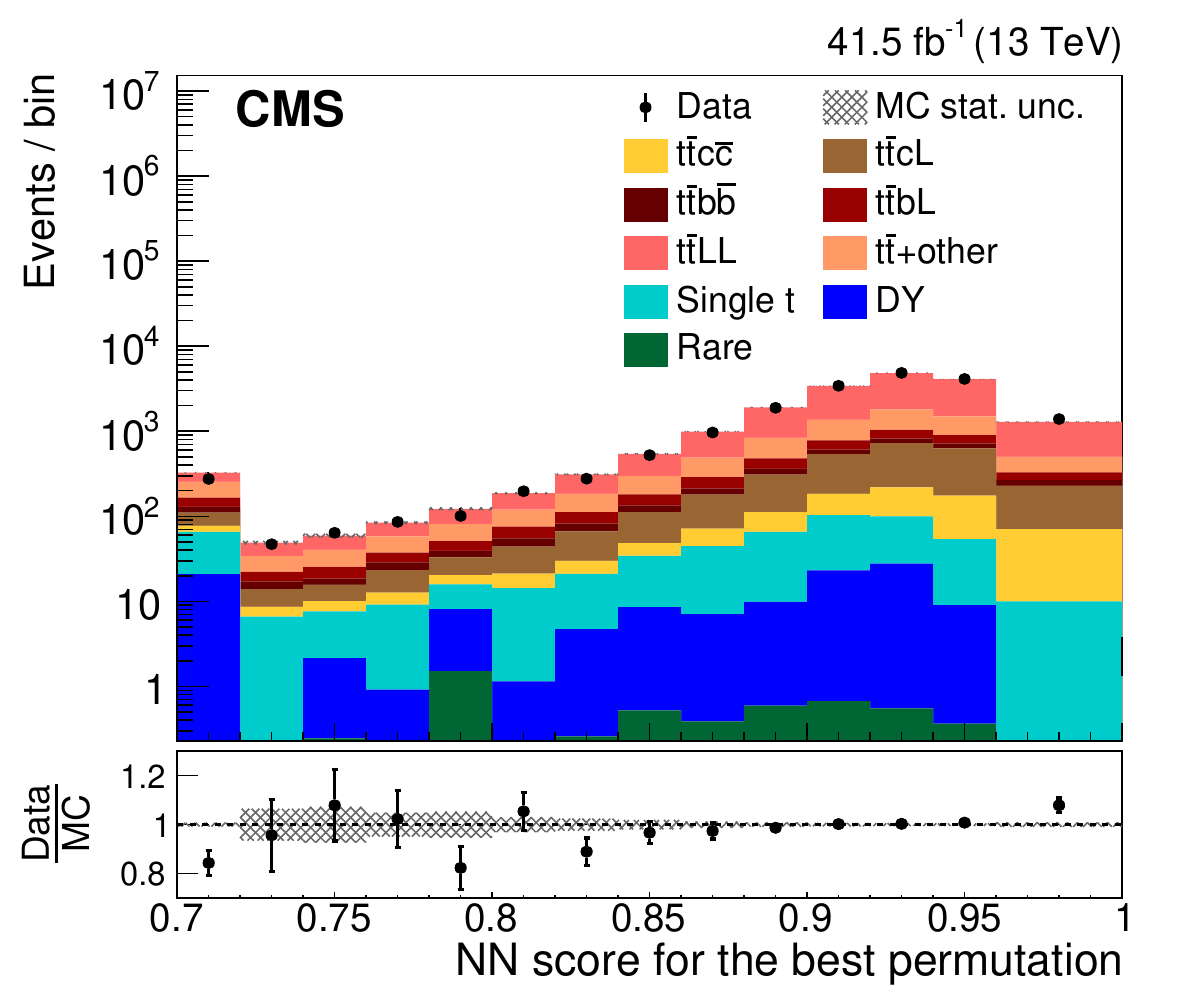}
\caption{\label{fig:DataToMCMatchingNN}
Comparison between data (points) and simulated predictions (histograms) for the distribution of the NN score for the best permutations of jet-parton assignments found in each event. Underflow is included in the first bin. The distributions are found after the event selections outlined in Section~\ref{sec:selection}, but before fitting the predicted signal yields to the data. The lower panel shows the ratio of the yields in data to those predicted in simulations. The vertical bars represent the statistical uncertainties in data, while the hatched bands show the statistical uncertainty in the simulated predictions.}
\end{figure}

The main objective is to identify additional HF jets in the event. In the assignment of {\PQb} jets from the $\ttbar$ decays, it does not matter which {\PQb} jet is matched to the decay of the top quark or antiquark. Any permutation for which these {\PQb} jets are reversed can still be considered appropriate for this measurement. If at least one additional {\PQb} or {\PQc} jet is present, a correct permutation has to identify these as the first or second additional jets. The NN is also trained to choose a permutation in which the first additional jet has a larger {\PQb} tagging discriminator value than the second additional jet. With these considerations in mind, three output classes are defined. One of the NN outputs denotes the probability for a given permutation to correspond to the correct jet-parton assignment ($P^{+}$). Another output class refers to those permutations for which the additional jets are correctly matched, but the {\PQb} jets from the \ttbar decays are reversed as explained above ($P^{\times}$). The third output represents all permutations for which the matching is wrong ($P^{-}$), meaning that either at least one of the {\PQb} jets from the \ttbar decays is not correctly matched, or an additional HF jet is found but is not identified as either the first or the second additional jet. The best jet-parton assignment is then identified by selecting the permutation with the highest value of:
\begin{linenomath}
\begin{equation}
\max\left( \frac{P^{+}}{P^{+}+P^{-}} ,\frac{P^{\times}}{P^{\times}+P^{-}} \right).
\label{eq:BestPermutation}
\end{equation}
\end{linenomath}
The NN is trained using the \textsc{Keras} deep learning library \cite{chollet2015keras}, interfaced to \textsc{TensorFlow} \cite{tensorflow2015-whitepaper} as a back end. Its architecture is composed of two fully connected hidden layers with 50 neurons each, and with a rectified linear unit activation function. The training is performed using an independent data set of simulated $\ttbar$ events that pass all the event selections outlined in Section~\ref{sec:selection}, except for the {\PQb} tagging requirement.
Only events for which the two generator-level {\PQb} quarks from top quark decays lie within $\Delta\text{R} <$ 0.3 of a reconstructed {\PQb} jet are used for the training. These constitute $\approx$76\% of the simulated $\ttbar$ events with at least two additional jets. Of those, the NN correctly identifies the two additional {\PQc} ({\PQb}) jets in 50 (30)\% of the cases for $\ttcc$ ($\ttbb$) events. For $\ttbb$ events, the matching is more challenging because the HF tagging information cannot help in separating additional {\PQb} jets from those originating from top quark decays. A comparison between data and simulations of the NN score for the best permutation in each event is shown in Fig.~\ref{fig:DataToMCMatchingNN}, indicating good agreement between the two.

\section{Charm jet identification and calibration}\label{sec:ctagCalib}

The DeepCSV HF tagging algorithm has a multiclass output structure that predicts the probabilities for each jet to contain a single {\PQb} hadron ($P(\PQb)$), two {\PQb} hadrons ($P(\PQb\PQb)$), one or more {\PQc} hadrons ($P(\PQc)$), or no {\PQb} or {\PQc} hadrons ($P\text{({\PQu}{\PQd}{\PQs}{\Pg})}$). The {\PQb} tagger used throughout this analysis is constructed from the sum $P(\PQb)+P(\PQb\PQb)$. However, different combinations of output values provide different types of discrimination. Since the displacements of tracks and secondary vertices for {\PQc} jets are on average smaller than those for {\PQb} jets and larger than those for LF jets, a charm jet identification algorithm uses a combination of two discriminators. The first is used to distinguish {\PQc} jets from LF jets (CvsL) and the second separates {\PQc} jets from {\PQb} jets (CvsB). The CvsL and CvsB discriminators are defined from the multiclass output structure of the DeepCSV algorithm as:
\begin{linenomath}
\begin{equation}
  \begin{aligned}
\text{CvsL} &= \frac{P(\PQc)}{P(\PQc) + P\text{({\PQu}{\PQd}{\PQs}{\Pg})}}, \\
\text{CvsB} &= \frac{P(\PQc)}{P(\PQc) + P(\PQb) + P(\PQb\PQb)}.
  \end{aligned}
\label{eq:CTagDefinition}
\end{equation}
\end{linenomath}

The {\PQc} tagging discriminators require calibration with the data, given that these algorithms are trained on simulated events and are therefore prone to mismodelling effects in the input variables. The observed initial discrepancies between the data and the simulated predictions can be as large as 50\%, as can be seen from the distributions before calibration in \appen\ref{app:beforecTagCalib}. In order to use these {\PQc} tagger discriminators for the fit of the template distributions (as discussed in Section~\ref{sec:Fit}), the shape of the two-dimensional (CvsL, CvsB) distribution is corrected to reproduce the distribution observed in data. To this end, a novel calibration technique is employed that uses three control regions selecting for semileptonic $\ttbar$, {\PW}$+\,${\PQc}, and DY$\,+\,$jets events, which are enriched in {\PQb}, {\PQc}, and LF jets, respectively~\cite{CMS-PAS-BTV-20-001}. By means of an iterative fit in these three control regions, a set of scale factors for each jet flavour is derived, as a function of both the CvsL and CvsB discriminator values of a given jet. The effectiveness of this calibration for different jet flavours has been validated in the three corresponding control regions outlined above, showing that the calibrated distributions in simulations indeed match those in data within the associated uncertainties. Additionally, the method shows no bias when applied to pseudo-data constructed from simulated events that have artificial scale factors applied to them. 

\begin{figure*}[htb!]
\centering
\includegraphics[width=.49\textwidth]{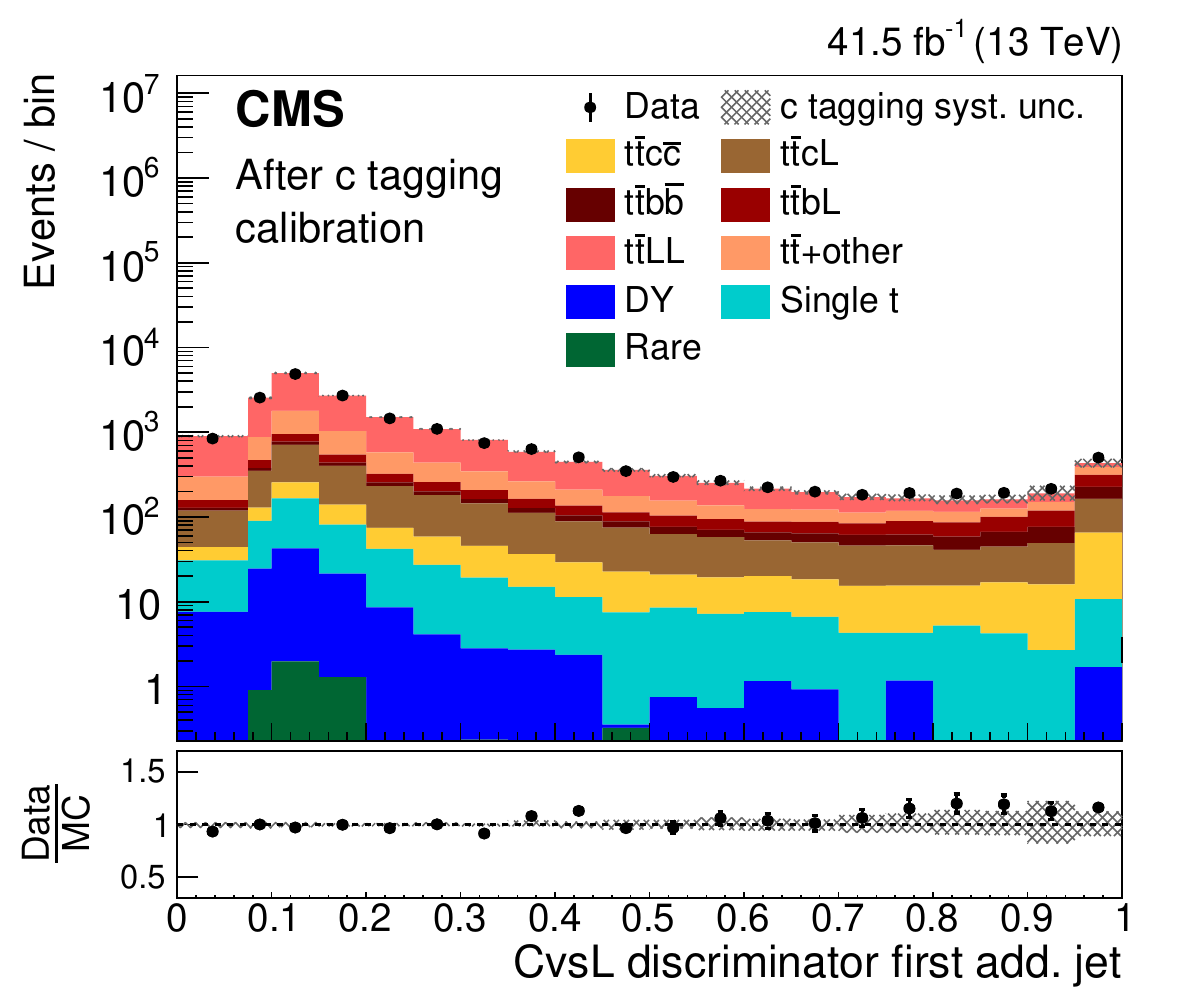}
\includegraphics[width=.49\textwidth]{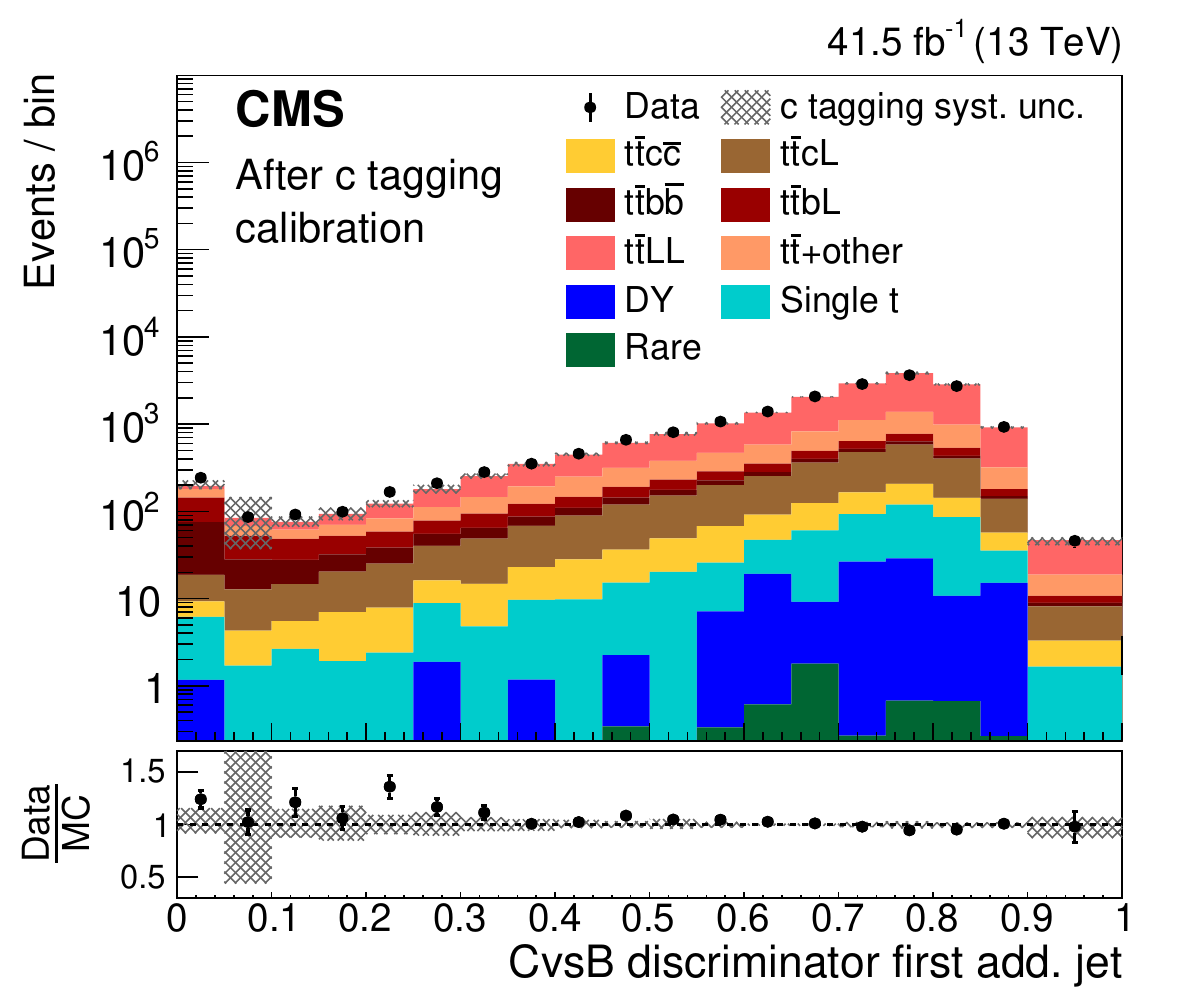}
\includegraphics[width=.49\textwidth]{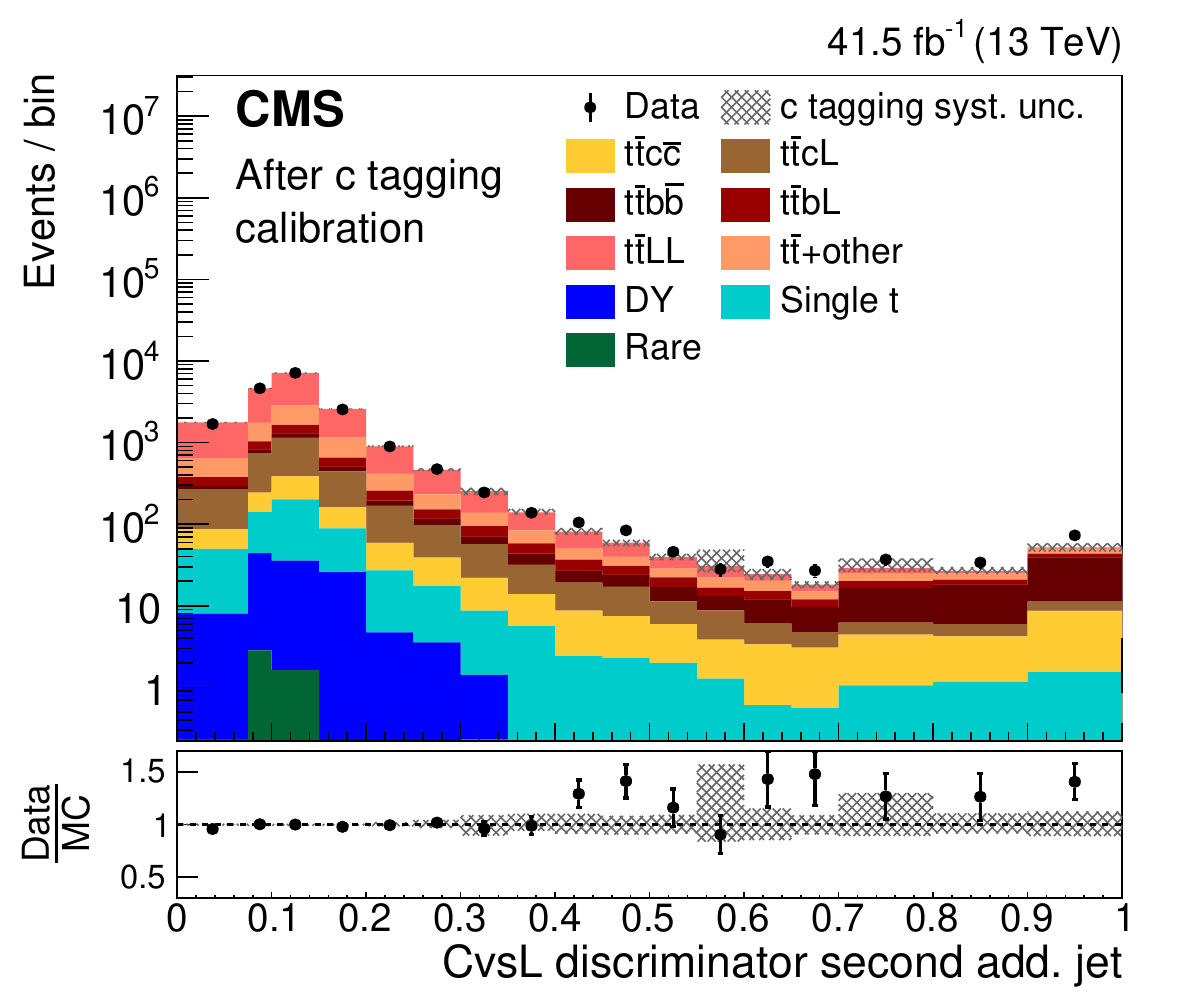}
\includegraphics[width=.49\textwidth]{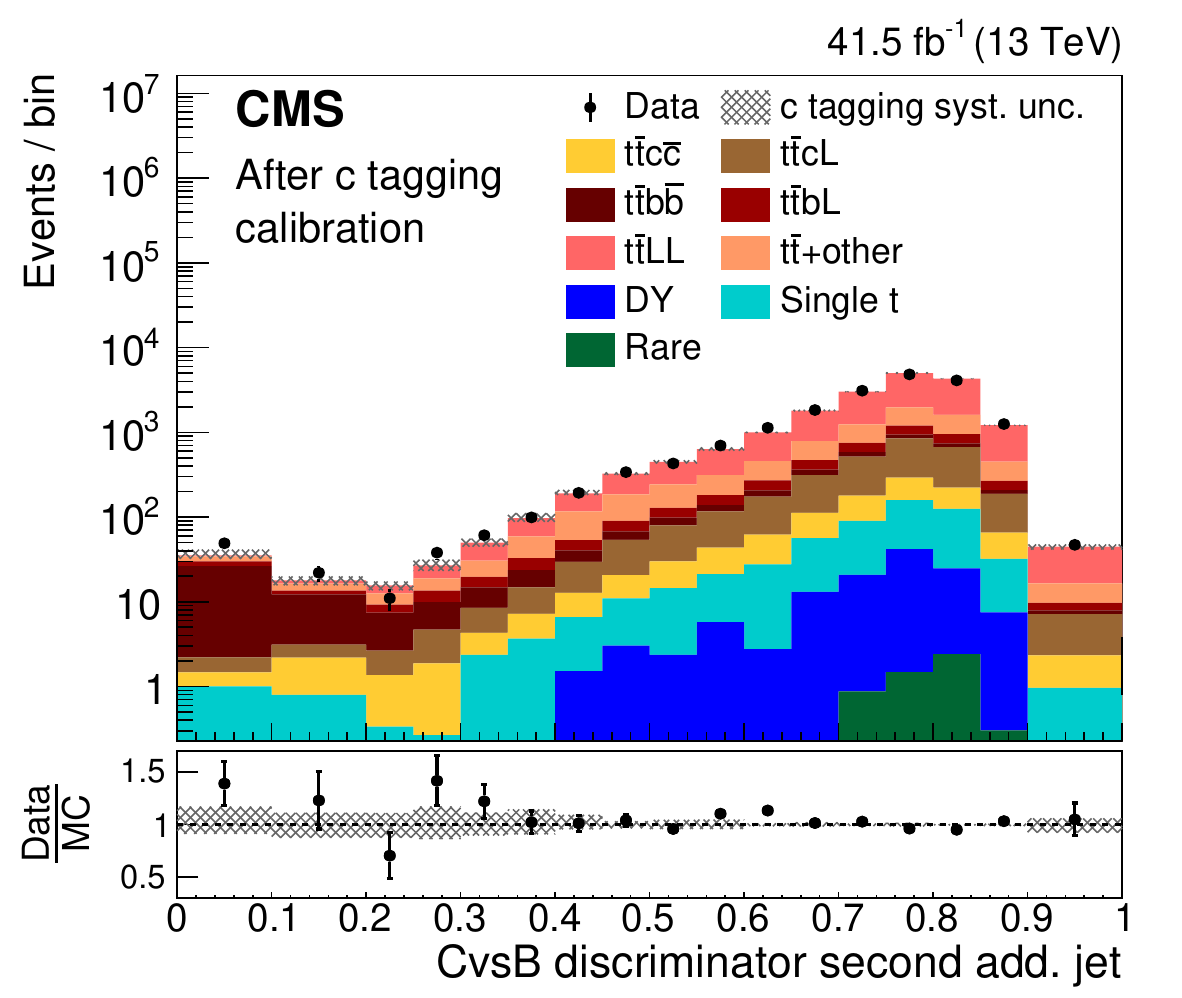}
\caption{\label{fig:CTagDilepton}
Comparison between data (points) and simulated predictions (histograms) for the CvsL (left column) and CvsB (right column) {\PQc} tagging discriminator distributions of the first (upper row) and second (lower row) additional jet after applying the {\PQc} tagging calibration. The distributions are found after the event selections outlined in Section~\ref{sec:selection}, but before fitting the predicted signal yields to the data. The lower panels show the ratio of the yields in data to those predicted in simulations. The vertical bars represent the statistical uncertainties in data, while the hatched bands show the systematic uncertainty from the {\PQc} tagging calibration only.}
\end{figure*}

After applying this calibration to the simulated events, the distributions of the CvsL and CvsB discriminators provide a good description of the data, as shown in Fig.~\ref{fig:CTagDilepton} for the first (upper row) and second (lower row) additional jet. The uncertainties related to this calibration are further discussed in Section~\ref{sec:uncertainties}.

\section{Fit to an event-based neural network discriminator}\label{sec:Fit}

The extraction of the $\ttcc$, $\ttbb$, and $\ttLF$ cross sections proceeds by means of a template fit to an observable that can distinguish among the different flavour categories. Since the differentiation relies mainly on the additional-jet flavour, the {\PQc} tagging discriminators of the first and second additional jets provide a natural choice for separating the different signals. The CvsL discriminator distinguishes the $\ttcc$ and $\ttbb$ from the $\ttLF$ events, whereas the CvsB discriminator provides additional information that can be used to distinguish between the $\ttcc$ and $\ttbb$ events. The flavour tagging information for the second additional jet allows the $\ttcL$ and $\ttbL$ processes to be identified. Additional information is extracted from two kinematic variables. The first is the angular separation $\Delta$R between the two additional jets. In the $\ttcc$ and $\ttbb$ processes, the additional jets arise predominantly from gluon splitting into $\PQc\PAQc$ and $\PQb\PAQb$ pairs, respectively, and are therefore expected to be more collimated. The second is the NN score for the best permutation (shown in Fig.~\ref{fig:DataToMCMatchingNN}), which is expected to be larger on average for $\ttLF$ events because the additional jets are well distinguished from the {\PQb} jets from top quark decays. This observable indirectly incorporates information on the event kinematic features through its input variables.

\begin{figure*}[ht!]
\centering
\includegraphics[width=1.\textwidth]{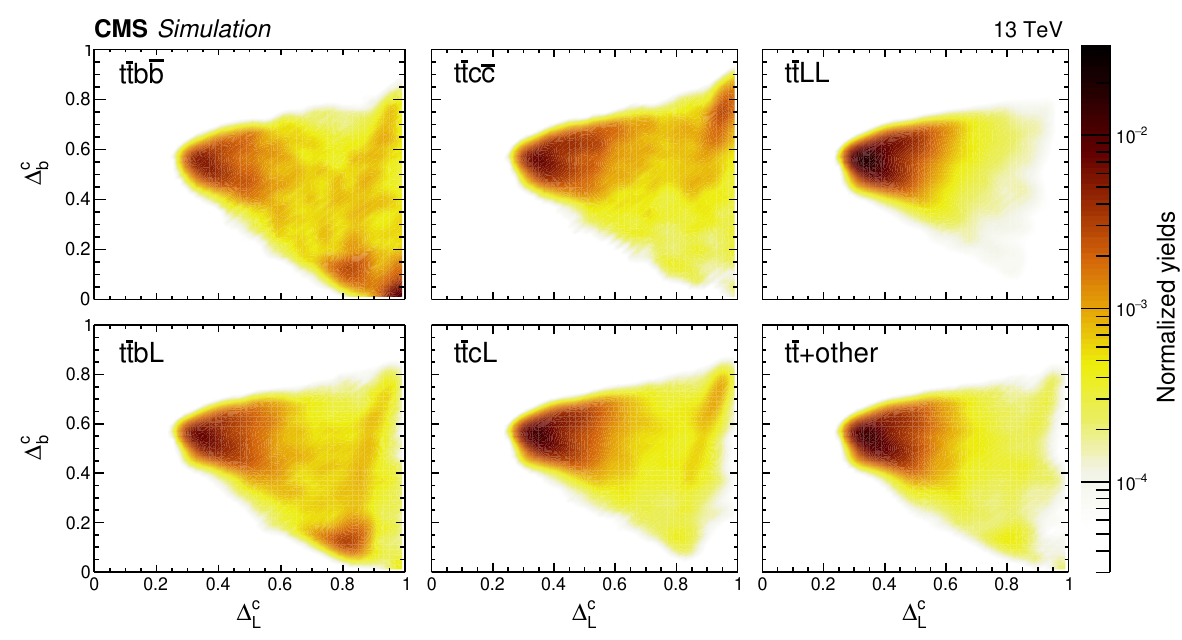}
\caption{\label{fig:FinalDiscr2D}
Normalized two-dimensional distributions of $\Delta_{\PQb}^{\PQc}$ vs. $\Delta_{\text{L}}^{\PQc}$ in simulated dileptonic $\ttbar$ events, for each of the event categories outlined in Section~\ref{sec:signal}. The colour scale on the right shows the normalized event yields.}
\end{figure*}

Using the six aforementioned observables, a NN is trained. Given the relatively small number of inputs, an architecture for the NN is chosen with one hidden layer that comprises 30 neurons with a rectified linear unit activation function. This NN predicts output probabilities for five output classes: $P$($\ttcc$), $P$($\ttcL$), $P$($\ttbb$), $P$($\ttbL$), and $P$($\ttLF$). To obtain the distributions that are used in the fit, these probabilities are projected onto a two-dimensional phase space spanned by two derived discriminators:
\begin{linenomath}
\begin{equation}
  \begin{aligned}
\Delta_{\PQb}^{\PQc} &= \frac{P(\ttcc) }{P(\ttcc) + P(\ttbb)},\\
\Delta_{\text{L}}^{\PQc} &=  \frac{P(\ttcc)}{P(\ttcc) + P(\ttLF)}.
\label{eq:Dbcdiscriminator}
  \end{aligned}
\end{equation}
\end{linenomath}
These discriminators can be interpreted as topology-specific {\PQc} tagger discriminators that augment the information on the jet flavour of the two additional jets with additional event kinematic features to optimally distinguish different signal categories. The two-dimensional distributions of these discriminators, normalized to unit area, are shown in Fig.~\ref{fig:FinalDiscr2D} for simulated dileptonic $\ttbar$ events. The different signal categories occupy different parts of this phase space, demonstrating that a fit of templates derived from these distributions to the data can be used to extract the $\ttcc$, $\ttbb$, and $\ttLF$ cross sections. It can be seen that the lower right corner of this phase space is almost exclusively populated with $\ttbb$ events, whereas the upper right corner is dominated by $\ttcc$ events.

Templates are constructed separately for each dilepton channel (ee, $\mu\mu$, and e$\mu$) and are fitted simultaneously to data by means of a binned maximum likelihood fit assuming Poisson statistics. The negative logarithm of the likelihood is minimized using the "Combine" framework developed for the combined Higgs boson measurements performed by ATLAS and CMS~\cite{Khachatryan:2016vau,Khachatryan:2014jba}. Uncertainties are included as nuisance parameters in the definition of the likelihood.

A first fit is performed to extract the absolute cross section ($\sigma$) of the $\ttcc$, $\ttbb$, and $\ttLF$ events in the fiducial phase space, where the expected yield in each bin is parametrized using the function:
\begin{linenomath}
\ifthenelse{\boolean{cms@external}}
{
\begin{equation}
\begin{aligned}
&F_{i}\left( \sigma_{\ttcc}, \, \sigma_{\ttbb}, \,  \sigma_{\ttLF} \right) = \\
&\Lumi^{\text{int}}   \left\{ \ 
\sigma_{\ttcc} \, \epsilon_{\ttcc} \left(  f^{\text{norm}}_{\ttcc,i} + \frac{N^{\text{MC}}_{\ttcL}}{N^{\text{MC}}_{\ttcc}} \, f^{\text{norm}}_{\ttcL,i}\right)  \right.   \\ 
& \qquad +  \sigma_{\ttbb} \,  \epsilon_{\ttbb} \left(  f^{\text{norm}}_{\ttbb,i} +  \frac{N^{\text{MC}}_{\ttbL}}{N^{\text{MC}}_{\ttbb}}  \, f^{\text{norm}}_{\ttbL,i}\right)  \\
& \qquad \left. + \; \sigma_{\ttLF} \,  \epsilon_{\ttLF} \left(  f^{\text{norm}}_{\ttLF,i} + \frac{N^{\text{MC}}_{\ttbar+\text{other}}}{N^{\text{MC}}_{\ttLF}} \,  f^{\text{norm}}_{\ttbar+\text{other},i} \right) \right\} \\
& + \Lumi^{\text{int}} \,   \sigma_{\text{bkg}} \, f^{\text{norm}}_{\text{bkg},i}   .
\end{aligned}
\label{eq:fitAbsoluteVisible} 
\end{equation}
}
{
\begin{equation}
\begin{aligned}
&F_{i}\left( \sigma_{\ttcc}, \, \sigma_{\ttbb}, \,  \sigma_{\ttLF} \right) = \\
&\Lumi^{\text{int}}   \left\{ \ 
\sigma_{\ttcc} \, \epsilon_{\ttcc} \left(  f^{\text{norm}}_{\ttcc,i} + \frac{N^{\text{MC}}_{\ttcL}}{N^{\text{MC}}_{\ttcc}} \, f^{\text{norm}}_{\ttcL,i}\right) +  \sigma_{\ttbb} \,  \epsilon_{\ttbb} \left(  f^{\text{norm}}_{\ttbb,i} +  \frac{N^{\text{MC}}_{\ttbL}}{N^{\text{MC}}_{\ttbb}}  \, f^{\text{norm}}_{\ttbL,i}\right) \right. \\
& \qquad +  \left. \, \sigma_{\ttLF} \,  \epsilon_{\ttLF} \left(  f^{\text{norm}}_{\ttLF,i} + \frac{N^{\text{MC}}_{\ttbar+\text{other}}}{N^{\text{MC}}_{\ttLF}} \,  f^{\text{norm}}_{\ttbar+\text{other},i} \right)
 + \,  \sigma_{\text{bkg}} \, f^{\text{norm}}_{\text{bkg},i}   \frac{}{} \right\}.
\end{aligned}
\label{eq:fitAbsoluteVisible}
\end{equation}
}
\end{linenomath}
Here, $\Lumi^{\text{int}}$ denotes the total integrated luminosity and $f^{\text{norm}}_{i}$ represents a given bin (with index~$i$) of the simulated two-dimensional template, normalized to unit area. The measurements in the reconstructed phase space are corrected to the fiducial phase space through an efficiency ($\epsilon$). To extract the result in the full phase space, an additional acceptance factor, $\mathcal{A}$, is applied to the extracted fiducial cross section to account for the difference in acceptance between the fiducial and full phase spaces. The efficiency and acceptance factors are summarized in Table~\ref{tab:VisiblePSEfficiencies}, and are calculated from simulations. The largest contribution to the acceptance can be attributed to the extrapolation from the dileptonic to the fully inclusive decays of the $\ttbar$ pairs. The remaining contributions are due to changes in kinematic requirements on the generator-level objects. As discussed in Section~\ref{sec:signal}, the $\ttcL$ and $\ttbL$ categories result from $\ttcc$ and $\ttbb$ events, respectively, where one of the additional HF jets is outside the acceptance or both are merged into one jet. Therefore, these components are scaled with the same factors as the $\ttcc$ and $\ttbb$ templates, respectively, such that the relative yield of $\ttcL$ ($\ttbL$) with respect to $\ttcc$ ($\ttbb$) events is fixed to that predicted in the simulations. The predicted yields from the simulations are denoted by $N^{\text{MC}}_{k}$, where $k$ denotes the signal process. The $\ttbar$+other component is scaled with the same factor as the $\ttLF$ component, motivated by their similar LF content. Uncertainties in the ratios of simulated yields are taken into account in the fit. The background processes are summed together into one template and their yield is fixed to the predictions from simulations, with uncertainties taken into account as discussed in Section~\ref{sec:uncertainties}. The sum of the cross sections for the background processes (obtained from simulations) is denoted as $\sigma_{\text{bkg}}$.

\begin{table}[ht!]
 \centering
  \topcaption{\label{tab:VisiblePSEfficiencies}
Selection efficiencies and acceptance factors for events in different signal categories. The values are obtained from simulated $\ttbar$ events.
}
  \begin{tabular}{ l  c c c c c }
  Event category & $\ttbb$  & $\ttbL$  & $\ttcc$ & $\ttcL$  & $\ttLF$  \\[0.08cm]
  \cline{1-6}\rule{0pt}{0.4cm}
  Efficiency $\epsilon$ (\%) & 12.0 & 8.5 & 6.8 & 5.6 & 4.5 \\
  Acceptance $\mathcal{A}$ (\%) & 2.9 & 2.5 & 2.0 & 2.0 & 2.3 \\
  \end{tabular}
\end{table}

A second binned maximum likelihood fit is performed assuming Poisson statistics to extract the ratios of the $\ttcc$ and $\ttbb$ cross sections to the overall inclusive $\ttjj$ cross section (denoted $R_{\PQc}$ and $R_{\PQb}$, respectively) in the fiducial phase space, with the expected yield in bin $i$ calculated using the function:
\begin{linenomath}
\ifthenelse{\boolean{cms@external}}
{
\begin{equation}
\begin{aligned}
&F_{i}\left( \sigma_{\ttjj}, R_{\PQc}, R_{\PQb} \right) =  \\
 &\Lumi^{\text{int}}  \sigma_{\ttjj}\left\{\ 
 R_{\PQc} \, \epsilon_{\ttcc} \left(  f^{\text{norm}}_{\ttcc,i} + \frac{N^{\text{MC}}_{\ttcL}}{N^{\text{MC}}_{\ttcc}} \, f^{\text{norm}}_{\ttcL,i}\right) \right. \\
& \qquad +  R_{\PQb} \, \epsilon_{\ttbb} \left(  f^{\text{norm}}_{\ttbb,i} +  \frac{N^{\text{MC}}_{\ttbL}}{N^{\text{MC}}_{\ttbb}} \, f^{\text{norm}}_{\ttbL,i}\right)   \\
& \qquad +  \left( 1- R_{\PQc}- R_{\PQc\text{L}} - R_{\PQb}- R_{\PQb\text{L}}  \right) \\
& \qquad \qquad \left. \epsilon_{\ttLF} \left( f^{\text{norm}}_{\ttLF,i} + \frac{N^{\text{MC}}_{\ttbar+\text{other}}}{N^{\text{MC}}_{\ttLF}} \, f^{\text{norm}}_{\ttbar+\text{other},i} \right)  \right\}    \\
& + \mathcal{L}^{\text{int}}  \, \sigma_{\text{bkg}} \, f^{\text{norm}}_{\text{bkg},i},   
\end{aligned}
\label{eq:fitRatioVisible}
\end{equation}
with
\begin{equation}
  \begin{cases}
 R_{\PQc\text{L}} =  \, R_{\PQc} \left( \frac{N^{\text{MC}}_{\ttcL} \, \epsilon_{\ttcc}}{N^{\text{MC}}_{\ttcc} \, \epsilon_{\ttcL}} \right), \\
R_{\PQb\text{L}} = \, R_{\PQb} \left( \frac{N^{\text{MC}}_{\ttbL} \, \epsilon_{\ttbb}}{N^{\text{MC}}_{\ttbb} \, \epsilon_{\ttbL}} \right). 
\end{cases}
\label{eq:ratios}
\end{equation}
}
{
  \begin{equation}
    \begin{aligned}
&F_{i}\left( \sigma_{\ttjj}, R_{\PQc}, R_{\PQb} \right) =  \\
 &\Lumi^{\text{int}}  \sigma_{\ttjj}\left\{\ 
 R_{\PQc} \, \epsilon_{\ttcc} \left(  f^{\text{norm}}_{\ttcc,i} + \frac{N^{\text{MC}}_{\ttcL}}{N^{\text{MC}}_{\ttcc}} \, f^{\text{norm}}_{\ttcL,i}\right)  +  R_{\PQb} \, \epsilon_{\ttbb} \left(  f^{\text{norm}}_{\ttbb,i} +  \frac{N^{\text{MC}}_{\ttbL}}{N^{\text{MC}}_{\ttbb}} \, f^{\text{norm}}_{\ttbL,i}\right) \right. \\
& \qquad \qquad + \left. \left( 1- R_{\PQc}- R_{\PQc\text{L}} - R_{\PQb} - R_{\PQb\text{L}}  \right)  \epsilon_{\ttLF} \left( f^{\text{norm}}_{\ttLF,i} + \frac{N^{\text{MC}}_{\ttbar+\text{other}}}{N^{\text{MC}}_{\ttLF}} \, f^{\text{norm}}_{\ttbar+\text{other},i} \right)  \right\}    \\
& + \mathcal{L}^{\text{int}}  \, \sigma_{\text{bkg}} \, f^{\text{norm}}_{\text{bkg},i},   \\
    \end{aligned}
    \label{eq:fitRatioVisible}
  \end{equation}
with
\begin{equation}
  \begin{cases}
 R_{\PQc\text{L}} =  \, R_{\PQc} \left( \frac{N^{\text{MC}}_{\ttcL} \, \epsilon_{\ttcc}}{N^{\text{MC}}_{\ttcc} \, \epsilon_{\ttcL}} \right), \\
R_{\PQb\text{L}} = \, R_{\PQb} \left( \frac{N^{\text{MC}}_{\ttbL} \, \epsilon_{\ttbb}}{N^{\text{MC}}_{\ttbb} \, \epsilon_{\ttbL}} \right). 
\end{cases}
\label{eq:ratios}
\end{equation}
}
\end{linenomath}
The parameters $R_{\PQc\text{L}}$ and $R_{\PQb\text{L}}$ are used to denote, respectively, the ratios of the $\ttcL$ and $\ttbL$ cross sections to the inclusive $\ttjj$ cross section, and are defined as a function of $R_{\PQc}$ and $R_{\PQb}$ in Eq.~\eqref{eq:ratios}.

\section{Systematic uncertainties}\label{sec:uncertainties}

This section summarizes the systematic uncertainties related to the extraction of the $\ttcc$, $\ttbb$, and $\ttLF$ cross sections (and the ratios $R_{\PQc}$ and $R_{\PQb}$), as well as corrections applied to the simulated events to account for differences with respect to data. Systematic uncertainties are treated as nuisance parameters in the template fit to data that can affect both the shapes of the templates and the yields of the signal and background processes. A smoothing procedure~\cite{Savitsky} is applied to the templates that describe the uncertainty variations affecting the template shape. The sources of systematic uncertainties are subdivided into experimental and theoretical components and are discussed below.

\textit{Experimental uncertainties}: These uncertainties affect both the shape and normalization of the templates. The jet energy resolution is known to be worse in data than in simulations, and a corresponding additional smearing is applied to the simulated jet energies~\cite{Khachatryan:2016kdb}. Systematic uncertainties are estimated by varying the smearing of the jet energy within its uncertainties in the calculation of the cross sections. Similarly, we take into account corrections and uncertainties from observed differences in the jet energy scale. These corrections are evaluated and applied in different regions of jet $\pt$ and $\abs{\eta}$.
Observed differences in electron and muon identification, isolation, reconstruction, and trigger efficiencies between data and simulations are taken into account through $\pt$- and $\eta$-dependent scale factors, with the corresponding uncertainties accounted for.
The distribution of the number of pileup collisions in simulated events is reweighted to match the distribution observed in data, using an inelastic $\Pp\Pp$ cross section of 69.2\unit{mb}~\cite{Sirunyan:2018nqx}. An uncertainty related to this correction is applied by varying this inelastic cross section by $\pm$ 4.6\%.
An uncertainty of 2.3\%~\cite{CMS-PAS-LUM-17-004} in the total integrated luminosity is also taken into account. The scale factors extracted from the {\PQc} tagging calibration are applied to the simulated events, and corresponding uncertainties are considered. Uncertainties related to this calibration are found to be dominated by the choice of $\mu_{\text{R}}$ and $\mu_{\text{F}}$ in the {\PW}$+\,${\PQc} and DY$\,+\,$jets control regions, affecting scale factors for {\PQc} and LF jets, respectively, and by PS uncertainties in semileptonic $\ttbar$ events that affect scale factors for {\PQb} jets, together with a significant contribution from statistical uncertainties for all jet flavours. Most of the theoretical and experimental sources of uncertainty are in common between the control regions in which the {\PQc} tagging calibration is derived and the $\ttbar$ dileptonic signal region considered in this analysis. In such cases, the common uncertainties are considered fully correlated and evaluated simultaneously. 

\textit{Theoretical uncertainties}: In the ME calculation, the choice of $\mu_{\text{R}}$ and $\mu_{\text{F}}$ can have an impact on the kinematic distributions of the final-state objects. Uncertainties in these scales are taken into account by rescaling $\mu_\text{F}$ and $\mu_\text{R}$ up or down by a factor of two at the ME level~\cite{Cacciari:2003fi,Catani:2003zt}. The choice was made not to include the difference between \PYTHIA and alternative PS simulations as a systematic uncertainty in this analysis. Instead a variety of parameters in \PYTHIA, sensitive to the PS and hadronization, are consistently varied to assess the uncertainty in an unambiguous way. In the PS, the uncertainty in the value of the strong coupling constant (\alpS) evaluated at $m_{\PZ}$ is taken into account by varying the renormalization scale of QCD emissions in the initial- and final-state radiation up and down by a factor of two. Similarly, uncertainties in the momentum transfer from {\PQb} quarks to {\PQb} hadrons ({\PQb} fragmentation) in the PS have been included. The {\PQb} fragmentation in \PYTHIA is parametrized using a Bowler--Lund model~\cite{Bowler:1981sb,Andersson:1983ia,Sjostrand:1984ic}, with uncertainties calculated by tuning the internal parameters of this model to measurements from the ALEPH~\cite{Heister:2001jg}, DELPHI~\cite{DELPHI:2011aa}, OPAL~\cite{Abbiendi:2002vt} and SLD~\cite{Abe:2002iq} experiments. In practice, it was found that this parametrization can be varied within its uncertainties by a reweighting at the generator level of the so-called "transfer function", $x_{\PQb} = \pt(\PQb\text{ hadron})/\pt(\PQb\text{ jet})$. No such detailed assessment of the {\PQc} fragmentation uncertainties has yet been performed, but a similar level of variation is observed when comparing the available experimental measurements~\cite{Barate:1999bg} with the default \PYTHIA tune used in this analysis~\cite{Skands:2014pea}. We verified that variations of the {\PQc} jet transfer function ($x_{\PQc} = \pt(\PQc\text{ hadron})/\pt(\PQc\text{ jet})$) induce similar variations of the {\PQc} jet \pt and {\PQc} tagging discriminator distributions as the analogous variations in {\PQb} fragmentation for {\PQb} jets, and found that these effects are comfortably covered by an uncertainty a factor of two larger than that for {\PQb} fragmentation. This uncertainty is modelled as an uncertainty in the $\ttcc$ and $\ttcL$ yields in the fit in the fiducial phase space, and an additional uncertainty in the acceptance correction from the full to the fiducial phase space. Uncertainties associated with the PDF, as well as with the value of \alpS in the PDF of the proton are considered, following the PDF4LHC prescription~\cite{Butterworth:2015oua}. For all of the aforementioned theoretical sources of uncertainty (except for the {\PQc} fragmentation uncertainty), the effects of these variations on the shape of the fitted templates are taken into account in the fit, whereas their impact on the signal yields is considered as an uncertainty in the theoretical prediction to which the measurement is compared in Section~\ref{sec:Results}. The residual theoretical uncertainty that enters the measured cross sections through the $\epsilon_i$ terms in Eqs.~\eqref{eq:fitAbsoluteVisible} and~\eqref{eq:fitRatioVisible} is accounted for by a separate theoretical uncertainty in the efficiency. The experimental uncertainties in the efficiency are already taken into account through the uncertainties in the normalization of the templates. The matching between the ME and PS is governed by a parameter called $h_{\text{damp}}$. The value of this parameter is varied according to $h_{\text{damp}} = \left( 1.379^{\, +0.926}_{\, -0.505} \right)  m_{\PQt}$~\cite{Sirunyan:2019dfx} in a separate simulation. Since the size of this simulated data set is insufficient to reliably estimate the effect on the shapes of the templates, this uncertainty is conservatively estimated through its effect on the overall yield. The remnants of the $\Pp\Pp$ collisions that do not take part in the hard scattering are referred to as the underlying event. Their kinematic distributions are tuned in the generators to match those observed in the data~\cite{Sirunyan:2019dfx}. The resulting parametrization of the \textsc{CP5} tune, used in this analysis, is varied within its uncertainties in separate simulations. Here too the effect of the uncertainty in the overall yield, rather than in the template shapes, is propagated to the measured cross sections. The effects of the theoretical uncertainties listed above on the fixed ratios of $\ttbL$ to $\ttbb$, $\ttcL$ to $\ttcc$, and $\ttbar$+other to $\ttLF$ yields, taken from simulations in Eqs.~\eqref{eq:fitAbsoluteVisible} and~\eqref{eq:fitRatioVisible}, are also included in the fit. An uncertainty of 25\% is assigned to the total cross section of all background processes, based on the precision of recent measurements of the dominant background processes~\cite{Sirunyan:2017jej,Sirunyan:2018rlu}. The statistical uncertainty due to the finite size of the simulated samples is taken into account in the fit. In extrapolating the results from the fiducial to the full phase space, some of these theoretical uncertainties also affect the acceptance for the different signal categories. The individual and combined impacts of different sources of uncertainty in the acceptance are summarized in Table~\ref{tab:AcceptanceUncertainty}.

\begin{table}[ht]
\centering
 \topcaption{
Sources of theoretical uncertainties in the acceptance, used to extrapolate the results from the fiducial to the full phase space, for different signal categories, together with their individual impact in percent. The last row of the table quotes the total relative uncertainty in the acceptance, calculated by adding in quadrature the effects from individual sources. A dash indicates that the uncertainty is not applicable for that signal category.
}
\begin{tabular}{ l c c c c c }
Sources & \multicolumn{5}{c}{Uncertainty in the acceptance (\%)} \\
& $\ttcc$ & $\ttcL$ & $\ttbb$ & $\ttbL$ & $\ttLF$  \\[0.08cm]
\hline

$\mu_{\text{R}}$ and $\mu_{\text{F}}$ &  1.0 & 0.8 & 0.4 & 0.6 & 0.5  \\
PS scale &  1.6 & 1.6 & 1.2 & 1.4 & 1.5  \\
PDF &  0.6 & 0.8 & 1.2 & 1.2 & 0.9  \\
Underlying event &  1.0 & 1.2 & 1.8 & 1.1 & 0.3  \\
ME-PS matching &  2.1 & 2.3 & 1.0 & 2.5 & 1.7  \\
{\PQb} fragmentation &  2.2 & 2.0 & 4.2 & 2.6 & 2.3  \\
{\PQc} fragmentation &  4.5 & 2.0 & -- & -- & --  \\[0.15cm]

Total &  5.9 & 4.3 & 5.0 & 4.2 & 3.4

 \end{tabular}
 \label{tab:AcceptanceUncertainty}
\end{table}

The individual impacts from each source of uncertainty in the cross sections (and ratios) for the fiducial phase space after the fit are summarized in Table~\ref{tab:SummarySystUnc}. The fitted nuisance parameters do not deviate significantly from their initial values and are not significantly constrained. No strong correlations between any of the nuisance parameters and the fitted cross sections or ratios are observed. The dominant systematic uncertainties are related to the {\PQc} tagging calibration, followed by jet energy scale and fragmentation uncertainties, as well as uncertainties related to the matching between ME and PS, and the choice of $\mu_\text{R}$ and $\mu_\text{F}$ scales in the ME calculation.

\begin{table*}[ht!]
\centering
\topcaption{\label{tab:SummarySystUnc}
Classes of systematic uncertainties in the measured parameters and their individual impact in percent after the fit for the fiducial phase space. The upper (lower) rows of the table list uncertainties related to the experimental conditions (theoretical modelling). For classes describing contributions from multiple nuisance parameters, the quoted numbers are obtained by adding the impacts from individual sources within that class in quadrature. The last row gives the overall systematic uncertainty in each quantity, which results from the nuisance parameter variations in the fit and is not the quadrature sum of the individual components.  
}

\begin{tabular}{ l   c  c  c c  c   }
\multirow{2}{*}{Sources} & \multicolumn{5}{c}{Systematic uncertainty (\%)} \\
&  $\Delta\sigma_{\ttcc}$ & $\Delta\sigma_{\ttbb}$& $\Delta\sigma_{\ttLF}$& $\Delta$$R_{\PQc}$& $\Delta$$R_{\PQb}$  \\[0.08cm]  
\hline

Jet energy scale &                                                 4.0 & 3.2 & 4.7 & 2.8 & 2.1		\\
Jet energy resolution &                                            2.3 & 1.0 & 0.9 & 2.5 & 1.3		\\
c tagging calibration &                                            7.0 & 3.2 & 2.5 & 7.3 & 3.5		\\
Lepton identification and isolation &                              0.8 & 1.0 & 1.3 & 0.6 & 0.3		\\
Trigger &                                                          2.0 & 2.0 & 2.0 & $< 0.1$ & $< 0.1$		\\
Pileup &                                                           0.3 & 0.2 & 0.3 & 0.5 & $< 0.1$		\\
Total integrated luminosity &                                      2.3 & 2.4 & 2.3 & $< 0.1$ & $< 0.1$		\\[0.20cm]
&&&&&\\
$\mu_{\text{R}}$ and $\mu_{\text{F}}$ scales in ME &               3.3 & 6.2 & 2.1 & 3.8 & 6.8		\\
PS scale &                                                         0.4 & 1.6 & 0.3 & 0.5 & 1.6		\\
PDF &                                                              0.3 & 0.1 & 0.1 & 0.2 & 0.1		\\
ME-PS matching &                                                   7.1 & 5.7 & 3.5 & 2.6 & 1.5		\\
Underlying event &                                                 1.9 & 2.3 & 1.1 & 0.5 & 0.9		\\
{\PQb} fragmentation &                                             0.4 & 1.9 & 0.8 & 0.3 & 2.4		\\
{\PQc} fragmentation &                                             4.6 & $< 0.1$ & $< 0.1$ & 3.9 & 0.7		\\
$\ttbL(\PQc\text{L})$/$\ttbb(\ccbar)$ and $\ttbar$+other/$\ttLF$ & 2.4 & 1.8 & 1.1 & 1.8 & 1.5		\\
Efficiency (theoretical) &                                         2.4 & 2.1 & 2.0 & $< 0.1$ & $< 0.1$		\\
Simulated sample size &                                            3.2 & 2.6 & 1.1 & 3.1 & 2.5		\\
Background normalization &                                         0.5 & 0.7 & 0.6 & 0.1 & 0.1		\\[0.20cm]

Total &  															13.7 & 11.4 & 8.2 & 10.9 & 9.2

\end{tabular} 
\end{table*}

\section{Results}\label{sec:Results}

\begin{figure}[ht!]
\centering
\includegraphics[width=.49\textwidth]{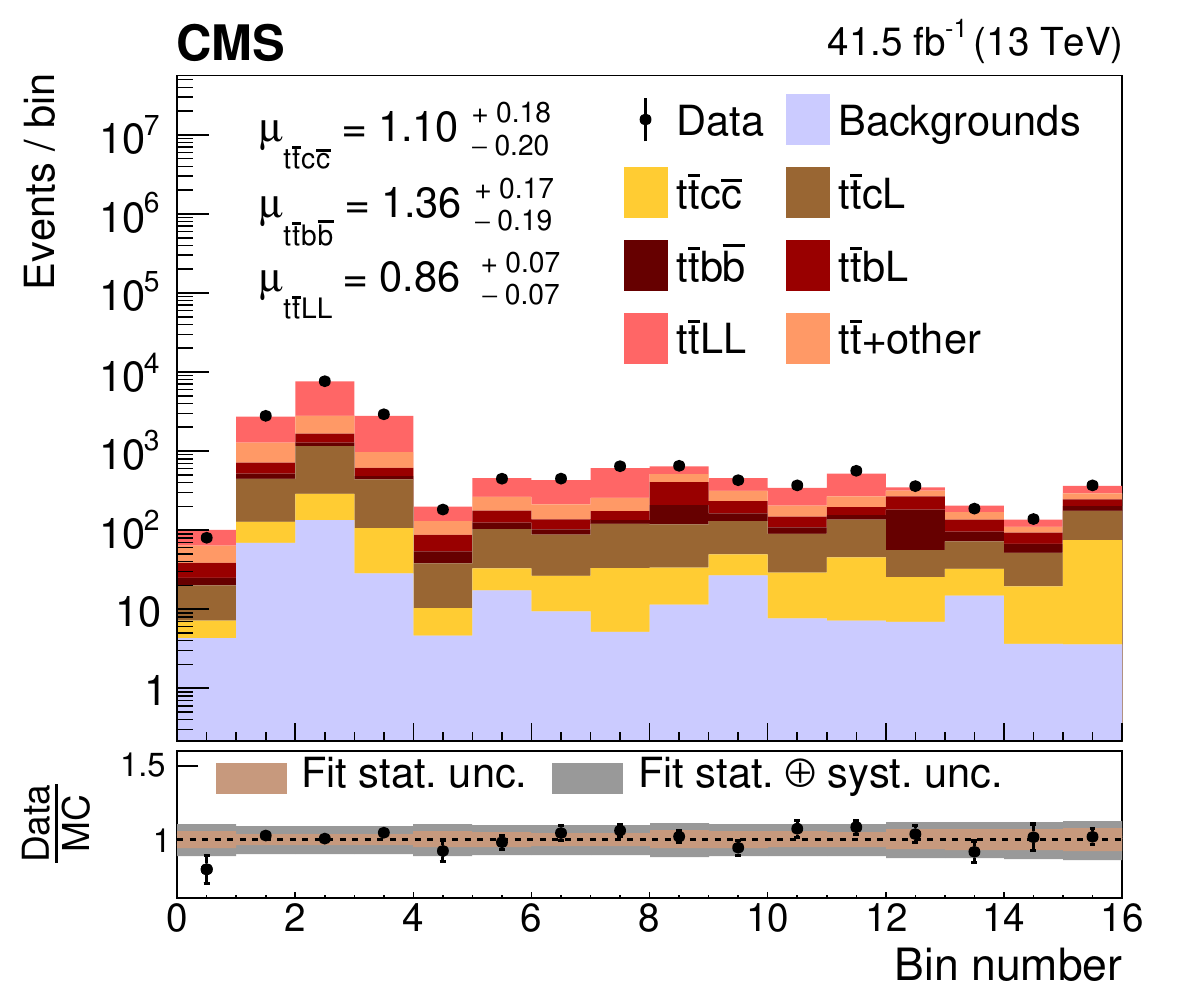}
\caption{\label{fig:PostFit}
A one-dimensional representation of the two-dimensional $\Delta_{\text{L}}^{\PQc}$ vs. $\Delta_{\PQb}^{\PQc}$ distributions, in the simulations (histograms) and in data (points), after normalizing the simulated templates according to the fitted cross sections. The lower panel shows the ratio of the yields in data to those predicted in the simulations. The brown and grey uncertainty bands denote, respectively, the statistical and total uncertainties from the fit. The factors ($\mu$) by which the templates of the different processes (using the \POWHEG ME generator) are scaled, are also displayed, together with their combined statistical and systematic uncertainties.}
\end{figure}

The binning of the two-dimensional $\Delta_{\text{L}}^{\PQc}$ vs. $\Delta_{\PQb}^{\PQc}$ distribution is chosen to be:
\begin{linenomath}
\ifthenelse{\boolean{cms@external}}
{
\begin{equation}  
\begin{aligned}
\Delta_{\text{L}}^{\PQc} &\otimes \Delta_{\PQb}^{\PQc} : \\
 [0,0.55,0.65,0.85,1.0] &\otimes [0,0.35,0.5,0.6,1.0].
\end{aligned}
\end{equation}
}
{
\begin{equation}
\Delta_{\text{L}}^{\PQc} \otimes \Delta_{\PQb}^{\PQc} :  [0,0.55,0.65,0.85,1.0] \otimes [0,0.35,0.5,0.6,1.0].
\end{equation}
}
\end{linenomath}
This gives a total of 16 bins with varying compositions of signal categories. From these 16 bins, a one-dimensional histogram is created in which the first four bins correspond to the first bin in $\Delta_{\text{L}}^{\PQc}$, with increasing values of $\Delta_{\PQb}^{\PQc}$, and then analogously for the remainder of the bins. This histogram is shown in Fig.~\ref{fig:PostFit} after normalizing the simulated templates according to the fitted cross sections. The factors by which the templates of the $\ttcc$, $\ttbb$, and $\ttLF$ processes (using the \POWHEG ME generator) are scaled to match the data are denoted $\mu_{\ttcc}$, $\mu_{\ttbb}$, and $\mu_{\ttLF}$, respectively. Their measured values from the fit are displayed in the top panel of Fig.~\ref{fig:PostFit}, together with their combined statistical and systematic uncertainties. An improved agreement between data and simulated predictions is also observed for the $\PQc$ tagging discriminators of the first and second additional jets after normalizing the simulated templates according to the fitted cross sections. This is demonstrated in \appen\ref{app:CtagDiscrAfterFitting} and can be directly compared to the agreement before the fit in Fig.~\ref{fig:CTagDilepton}.

The measured cross sections in the fiducial and full phase spaces, together with their statistical and systematic uncertainties, are summarized in Table~\ref{tab:SummaryResults} and compared to predictions using the \POWHEG and \MGvATNLO ME generators. Uncertainties in the measured values of the cross sections and ratios are determined from the points where the decrease in the logarithm of the profiled likelihood from the best-fit value intersects with 0.5. The inclusive $\ttcc$ cross section and the ratio $R_{\PQc}$ are measured here for the first time. They are in agreement with the predictions from both ME generators, within the uncertainties.

 \begin{table*}[ht!]
 \centering
  \topcaption{\label{tab:SummaryResults}
Measured parameter values in the fiducial (upper rows) and full (lower rows) phase spaces with their statistical and systematic uncertainties listed in that order. The last two columns display the expectations from the simulated $\ttbar$ samples using the \POWHEG or \MGvATNLO ME generators. The uncertainties quoted for these predictions include the contributions from the theoretical uncertainties listed in the lower rows of Table~\ref{tab:SummarySystUnc}, as well as the uncertainty in the $\ttbar$ cross section.
}
\begin{tabular}{ l   D{,}{\, \pm \,}{4.13}  D{,}{\, \pm \,}{4.4} D{,}{\, \pm \,}{4.4} }
   & \multicolumn{1}{c}{Result}  & \multicolumn{1}{c}{\text{\POWHEG}} &  \multicolumn{1}{c}{\text{\footnotesize \MGvATNLO}}   \\[0.08cm]
  \hline
   
 \multicolumn{4}{c}{Fiducial phase space}\\[0.08cm]
   $ \sigma_{\ttcc}$ [pb] & 0.207 , 0.025 \pm 0.027 \; & 0.187 , 0.038 & 0.189 , 0.032 \\
   $ \sigma_{\ttbb}$ [pb] & 0.132 , 0.010 \pm 0.015 \; & 0.097 , 0.021 & 0.101 , 0.023 \\
   $ \sigma_{\ttLF}$ [pb] & 5.15 , 0.12 \pm 0.41 \; & 5.95 , 1.02 & 6.32 , 0.94 \\
   $R_{\PQc}$ [\%] & 3.01 , 0.34 \pm 0.31 \; & 2.53 , 0.18 & 2.43 , 0.17 \\
   $R_{\PQb}$ [\%] & 1.93 , 0.15 \pm 0.18 \; & 1.31, 0.12 & 1.30 , 0.16 \\
   $$$$ \\
   \multicolumn{4}{c}{Full phase space}\\[0.08cm]
   $ \sigma_{\ttcc}$ [pb] & 10.1 , 1.2 \pm 1.4 \; & 9.1 , 1.8 & 8.9 , 1.5 \\
   $ \sigma_{\ttbb}$ [pb] & 4.54 , 0.35 \pm 0.56 \; & 3.34 , 0.72 & 3.39 , 0.66 \\
   $ \sigma_{\ttLF}$ [pb] & 220 , 5 \pm 19 \; & 255 , 43 & 261 , 37 \\
   $R_{\PQc}$ [\%] & 3.36 , 0.38 \pm 0.34 \; & 2.81 , 0.20 & 2.72 , 0.19 \\
   $R_{\PQb}$ [\%] & 1.51 , 0.11 \pm 0.16 \; & 1.03 , 0.08 & 1.03 , 0.09 \\

  \end{tabular}
\end{table*}

\begin{figure*}[htb!]
\centering
\includegraphics[width=.45\textwidth]{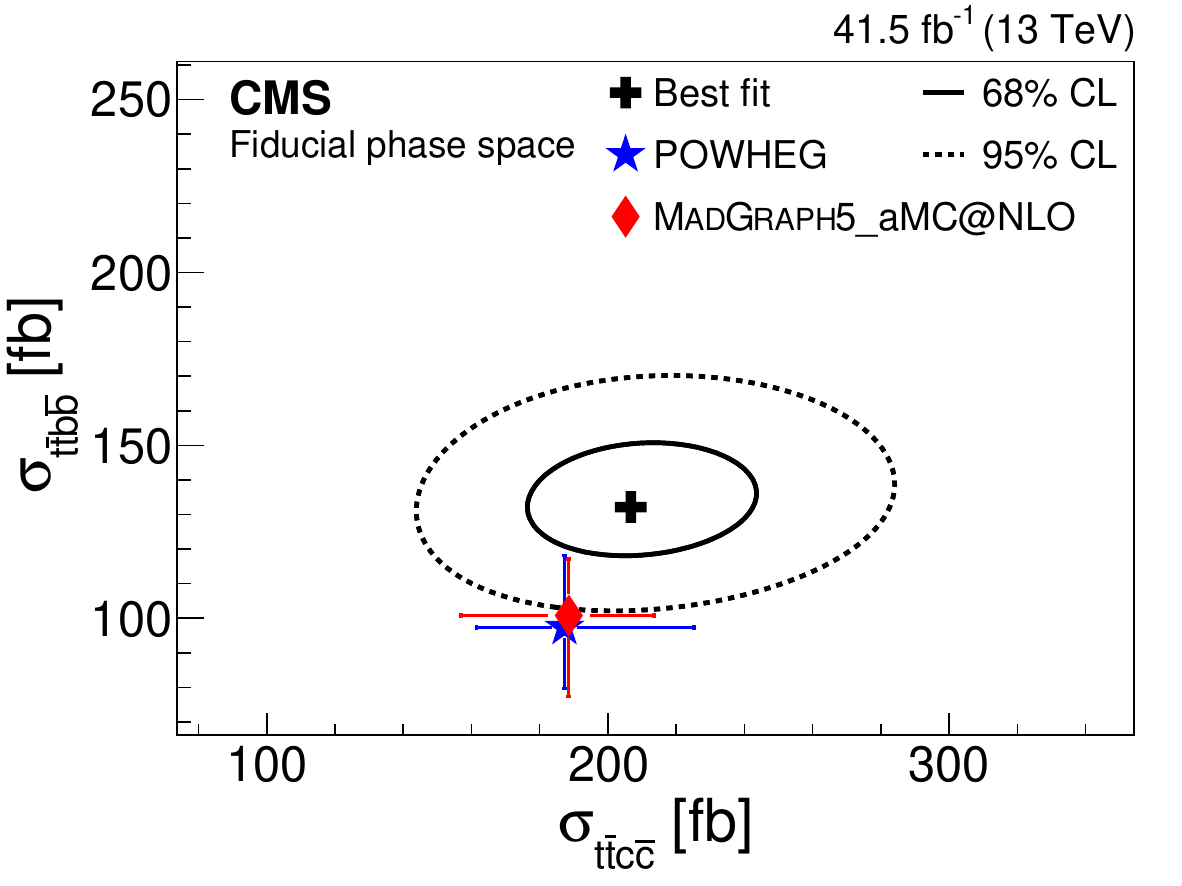}
\includegraphics[width=.45\textwidth]{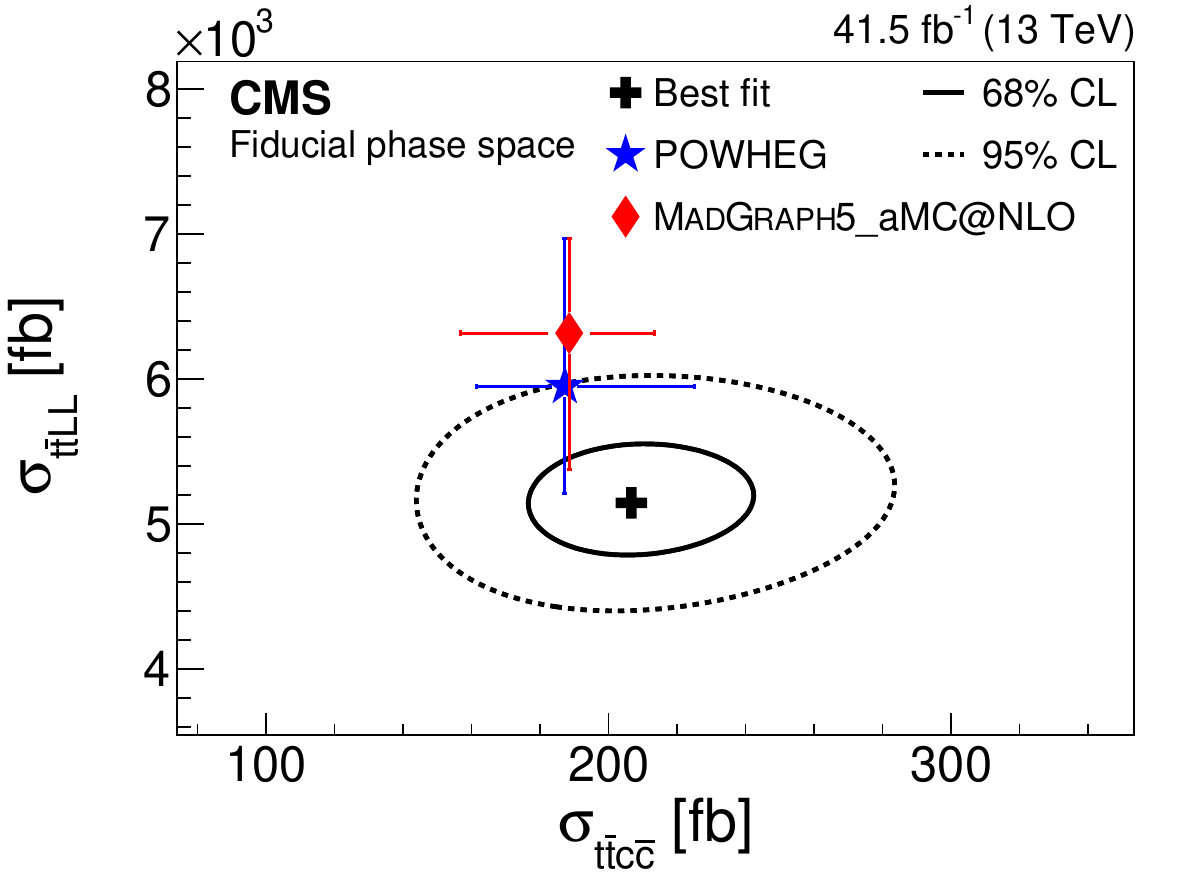}
\includegraphics[width=.45\textwidth]{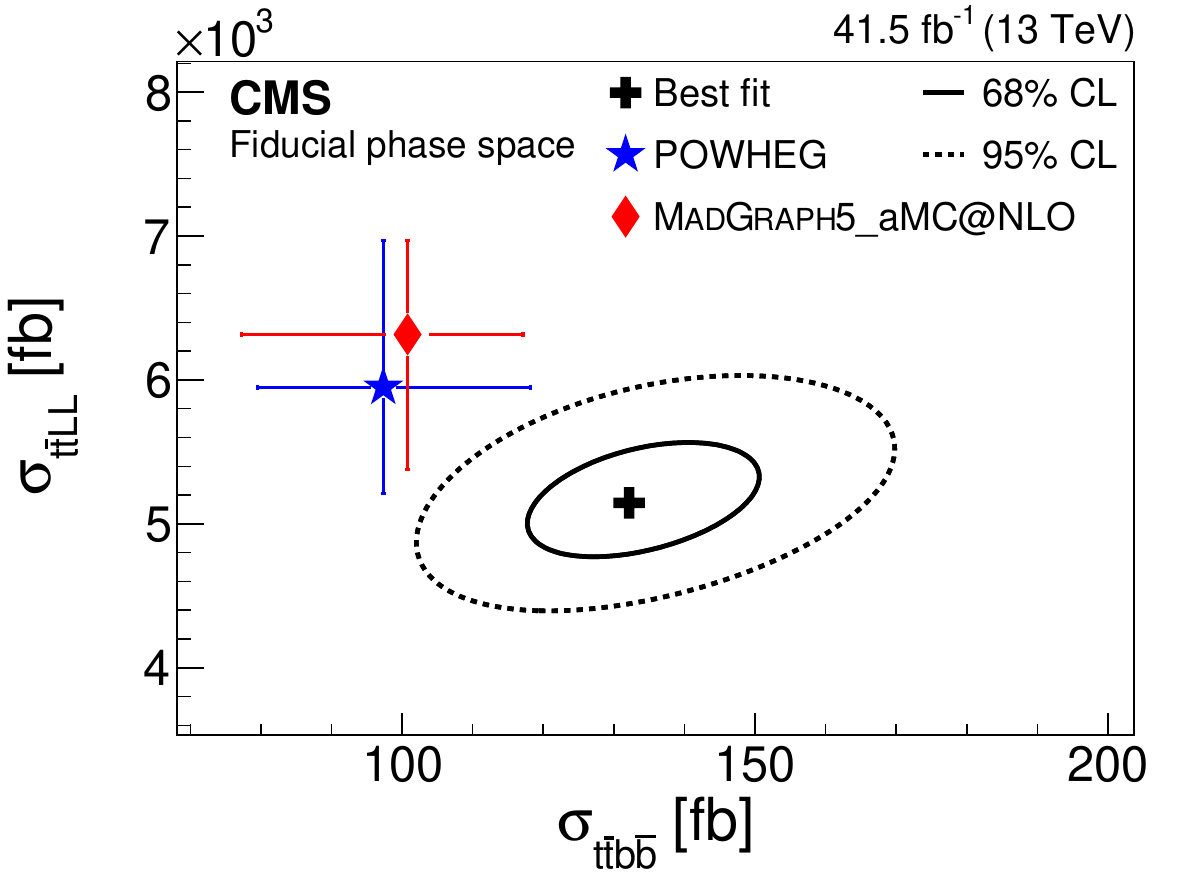}
\includegraphics[width=.45\textwidth]{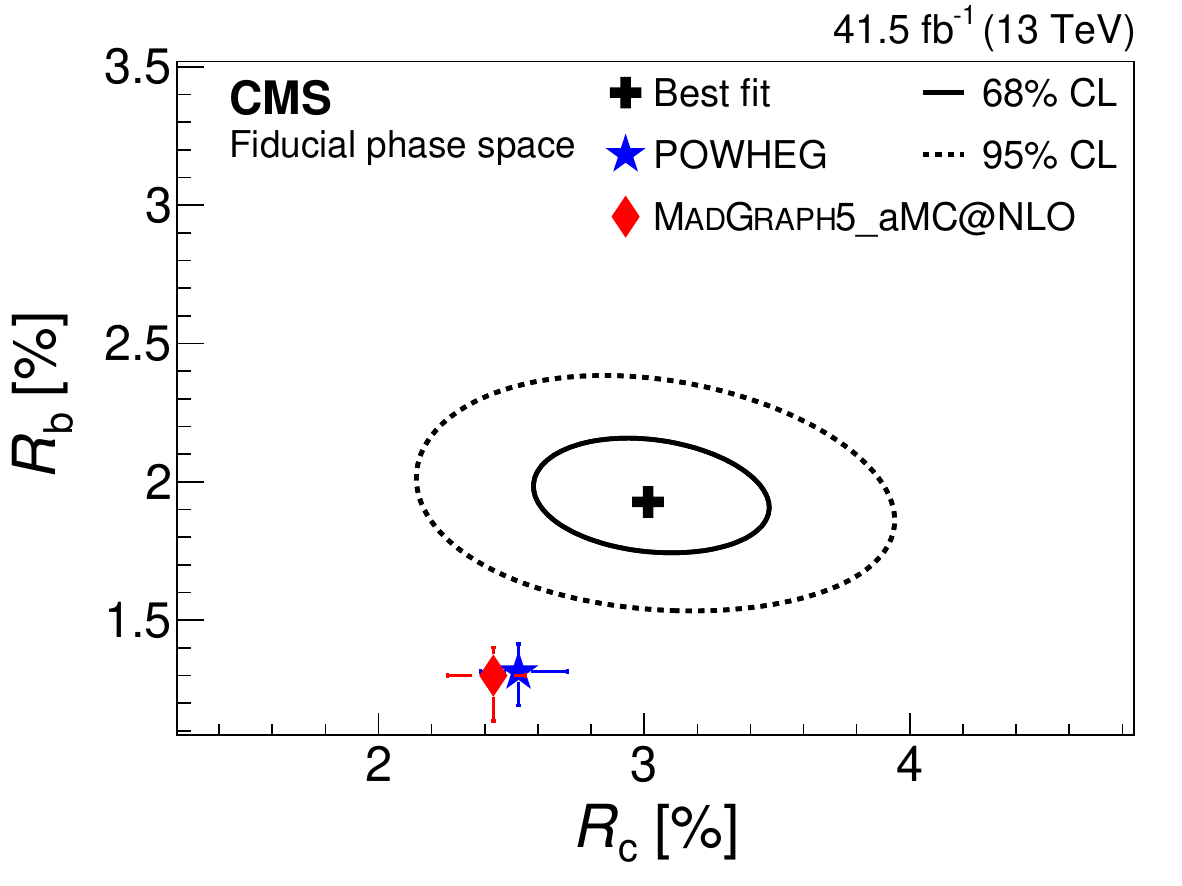}
\caption{\label{fig:likelihoodScans2D}
Results of the two-dimensional likelihood scans for several combinations of the parameters of interest in the fiducial phase space. The best-fit value (black cross) with the corresponding 68\% (full) and 95\% (dashed) confidence level (\CL) contours are shown, compared to the theoretical predictions using either the \POWHEG (blue star) or \MGvATNLO (red diamond) ME generators. Uncertainties in the theoretical predictions are displayed by the horizontal and vertical bars on the markers.}
\end{figure*}

In addition, two-dimensional likelihood scans are performed over different combinations of cross sections or ratios. These are shown in Fig.~\ref{fig:likelihoodScans2D} for the measurements in the fiducial phase space. The 68 and 95\% confidence level contours are shown, along with the predictions from the \POWHEG and \MGvATNLO ME generators. Agreement is observed at the level of one to two standard deviations between the measured values and simulated predictions for the $\ttcc$, $\ttbb$, and $\ttLF$ processes. The most significant tension is observed in the ratio $R_{\PQb}$, at the level of 2.5 standard deviations. Results for $\sigma_{\ttbb}$ and $R_{\PQb}$ are consistent with previous measurements targeting specifically this signature~\cite{Aad:2013tua,CMS:2014yxa,Aad:2015yja,Khachatryan:2015mva,Sirunyan:2017snr,Aaboud:2018eki,Sirunyan:2019jud,Sirunyan:2020kga} and show the same tendency to be slightly above the predictions from simulations.

\section{Summary}\label{sec:Conclusion}

The production of a top quark pair ($\ttbar$) in association with additional bottom or charm jets at the LHC provides challenges both in the theoretical modelling and experimental measurement of this process.
Whereas $\ttbar$ production with two additional bottom jets ($\ttbb$) has been measured by the ATLAS and CMS Collaborations at different centre-of-mass energies~\cite{Aad:2013tua,CMS:2014yxa,Aad:2015yja,Khachatryan:2015mva,Sirunyan:2017snr,Aaboud:2018eki,Sirunyan:2019jud,Sirunyan:2020kga}, this analysis presents the first measurement of the cross section for $\ttbar$ production with two additional charm jets ($\ttcc$). The analysis is conducted using data from proton-proton collisions recorded by the CMS experiment at a centre-of-mass energy of $13\TeV$, corresponding to an integrated luminosity of 41.5\fbinv. The measurement is performed in the dileptonic channel of the $\ttbar$ decays and relies on the use of recently developed charm jet identification algorithms ({\PQc} tagging). A template fitting method is used, based on the outputs of a neural network classifier trained to identify the signal categories defined by the flavour of the additional jets. This allows the simultaneous extraction of the cross section for the $\ttcc$, $\ttbb$, and $\ttbar$ with two additional light-flavour or gluon jets ($\ttLF$) processes. A novel multidimensional calibration of the shape of the {\PQc} tagging discriminator distributions is employed, such that this information can be reliably used in the neural network classifier.

The $\ttcc$ cross section is measured for the first time to be $0.207\pm 0.025\stat\pm 0.027\syst\unit{pb}$ in the fiducial phase space (matching closely the sensitive region of the detector) and $10.1\pm 1.2\stat\pm 1.4\syst\unit{pb}$ in the full phase space. The ratio of the $\ttcc$ to the inclusive $\ttbar$ + two jets cross sections is found to be $(3.01\pm 0.34\stat\pm 0.31\syst)\%$ in the fiducial phase space and $(3.36\pm 0.38\stat\pm 0.34\syst)\%$ in the full phase space. These results are compared with predictions from two different matrix element generators with next-to-leading order accuracy in quantum chromodynamics, in which an inclusive description of the \ttbar process, with up to two additional radiated hard gluons at the ME level, is interfaced with a parton shower simulation to generate the additional radiation. Agreement is observed at the level of one to two standard deviations between the measured values and simulated predictions for the $\ttcc$, $\ttbb$, and $\ttLF$ processes. The observed ratio of the $\ttbb$ to the inclusive $\ttbar$ + two jets cross sections exceeds the predictions by about 2.5 standard deviations, consistent with the tendency seen in previous measurements.

\begin{acknowledgments}
  We congratulate our colleagues in the CERN accelerator departments for the excellent performance of the LHC and thank the technical and administrative staffs at CERN and at other CMS institutes for their contributions to the success of the CMS effort. In addition, we gratefully acknowledge the computing centres and personnel of the Worldwide LHC Computing Grid for delivering so effectively the computing infrastructure essential to our analyses. Finally, we acknowledge the enduring support for the construction and operation of the LHC and the CMS detector provided by the following funding agencies: BMBWF and FWF (Austria); FNRS and FWO (Belgium); CNPq, CAPES, FAPERJ, FAPERGS, and FAPESP (Brazil); MES (Bulgaria); CERN; CAS, MoST, and NSFC (China); COLCIENCIAS (Colombia); MSES and CSF (Croatia); RIF (Cyprus); SENESCYT (Ecuador); MoER, ERC PUT and ERDF (Estonia); Academy of Finland, MEC, and HIP (Finland); CEA and CNRS/IN2P3 (France); BMBF, DFG, and HGF (Germany); GSRT (Greece); NKFIA (Hungary); DAE and DST (India); IPM (Iran); SFI (Ireland); INFN (Italy); MSIP and NRF (Republic of Korea); MES (Latvia); LAS (Lithuania); MOE and UM (Malaysia); BUAP, CINVESTAV, CONACYT, LNS, SEP, and UASLP-FAI (Mexico); MOS (Montenegro); MBIE (New Zealand); PAEC (Pakistan); MSHE and NSC (Poland); FCT (Portugal); JINR (Dubna); MON, RosAtom, RAS, RFBR, and NRC KI (Russia); MESTD (Serbia); SEIDI, CPAN, PCTI, and FEDER (Spain); MOSTR (Sri Lanka); Swiss Funding Agencies (Switzerland); MST (Taipei); ThEPCenter, IPST, STAR, and NSTDA (Thailand); TUBITAK and TAEK (Turkey); NASU (Ukraine); STFC (United Kingdom); DOE and NSF (USA).
   
  \hyphenation{Rachada-pisek} Individuals have received support from the Marie-Curie programme and the European Research Council and Horizon 2020 Grant, contract Nos.\ 675440, 724704, 752730, and 765710 (European Union); the Leventis Foundation; the A.P.\ Sloan Foundation; the Alexander von Humboldt Foundation; the Belgian Federal Science Policy Office; the Fonds pour la Formation \`a la Recherche dans l'Industrie et dans l'Agriculture (FRIA-Belgium); the Agentschap voor Innovatie door Wetenschap en Technologie (IWT-Belgium); the F.R.S.-FNRS and FWO (Belgium) under the ``Excellence of Science -- EOS" -- be.h project n.\ 30820817; the Beijing Municipal Science \& Technology Commission, No. Z191100007219010; the Ministry of Education, Youth and Sports (MEYS) of the Czech Republic; the Deutsche Forschungsgemeinschaft (DFG) under Germany's Excellence Strategy -- EXC 2121 ``Quantum Universe" -- 390833306; the Lend\"ulet (``Momentum") Programme and the J\'anos Bolyai Research Scholarship of the Hungarian Academy of Sciences, the New National Excellence Program \'UNKP, the NKFIA research grants 123842, 123959, 124845, 124850, 125105, 128713, 128786, and 129058 (Hungary); the Council of Science and Industrial Research, India; the HOMING PLUS programme of the Foundation for Polish Science, cofinanced from European Union, Regional Development Fund, the Mobility Plus programme of the Ministry of Science and Higher Education, the National Science Center (Poland), contracts Harmonia 2014/14/M/ST2/00428, Opus 2014/13/B/ST2/02543, 2014/15/B/ST2/03998, and 2015/19/B/ST2/02861, Sonata-bis 2012/07/E/ST2/01406; the National Priorities Research Program by Qatar National Research Fund; the Ministry of Science and Higher Education, project no. 0723-2020-0041 (Russia); the Tomsk Polytechnic University Competitiveness Enhancement Program; the Programa Estatal de Fomento de la Investigaci{\'o}n Cient{\'i}fica y T{\'e}cnica de Excelencia Mar\'{\i}a de Maeztu, grant MDM-2015-0509 and the Programa Severo Ochoa del Principado de Asturias; the Thalis and Aristeia programmes cofinanced by EU-ESF and the Greek NSRF; the Rachadapisek Sompot Fund for Postdoctoral Fellowship, Chulalongkorn University and the Chulalongkorn Academic into Its 2nd Century Project Advancement Project (Thailand); the Kavli Foundation; the Nvidia Corporation; the SuperMicro Corporation; the Welch Foundation, contract C-1845; and the Weston Havens Foundation (USA).
\end{acknowledgments}

\ifthenelse{\boolean{cms@external}}{}{\clearpage}
\bibliography{auto_generated} 

\appendix
\section{Charm tagging discriminators before calibration}\label{app:beforecTagCalib}
\begin{figure*}[htb!]
  \centering
  \includegraphics[width=.49\textwidth]{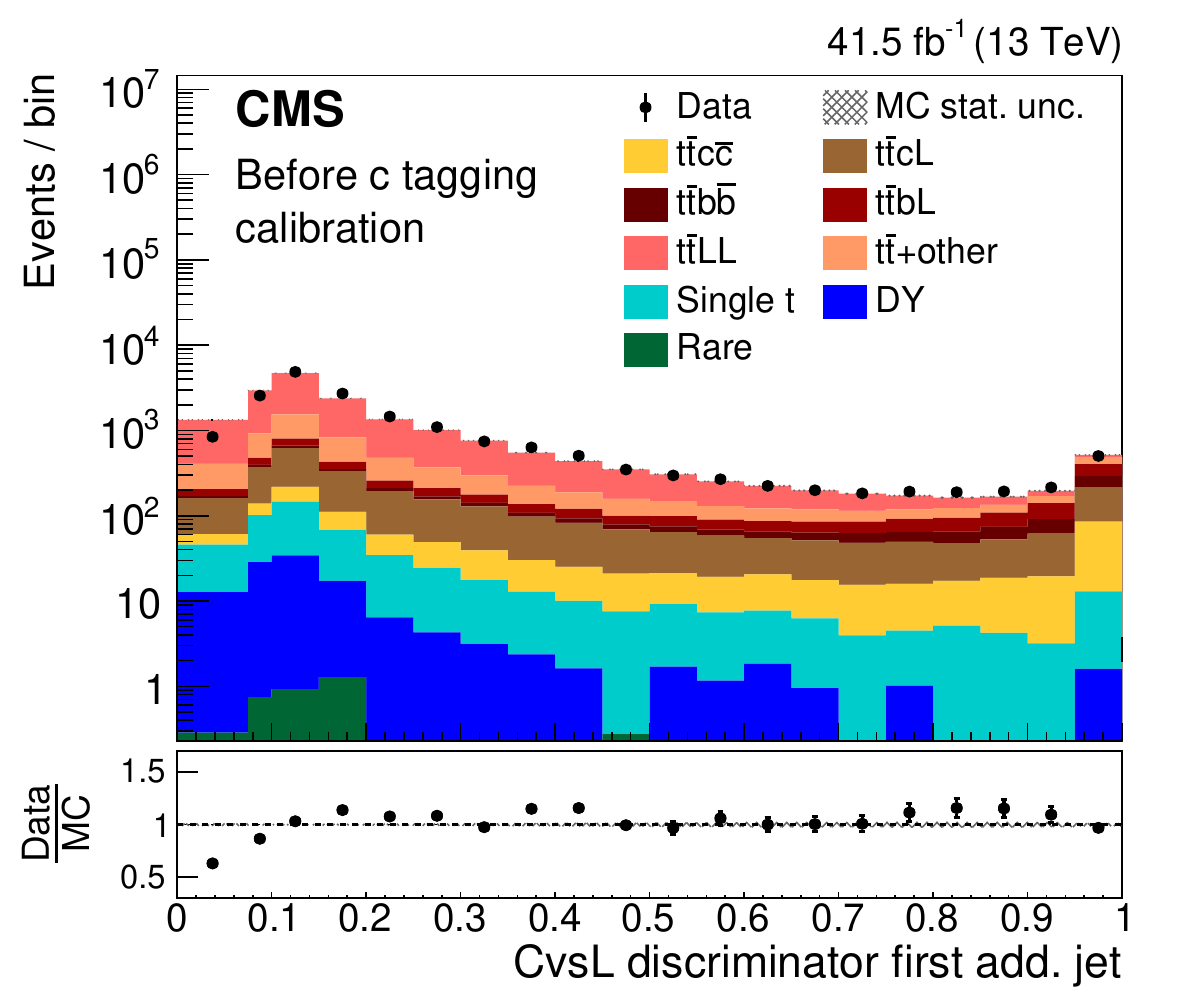}
  \includegraphics[width=.49\textwidth]{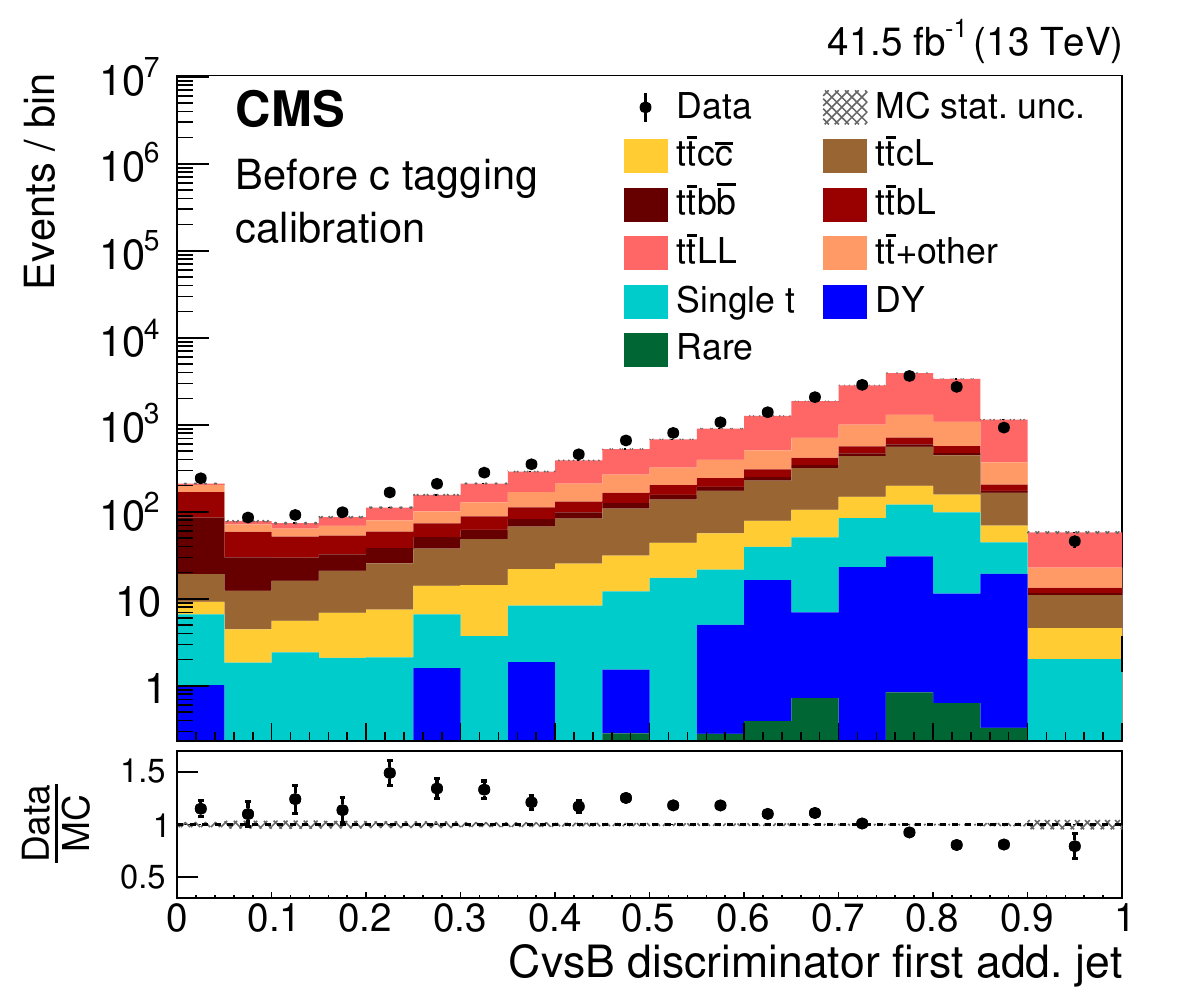}
  \includegraphics[width=.49\textwidth]{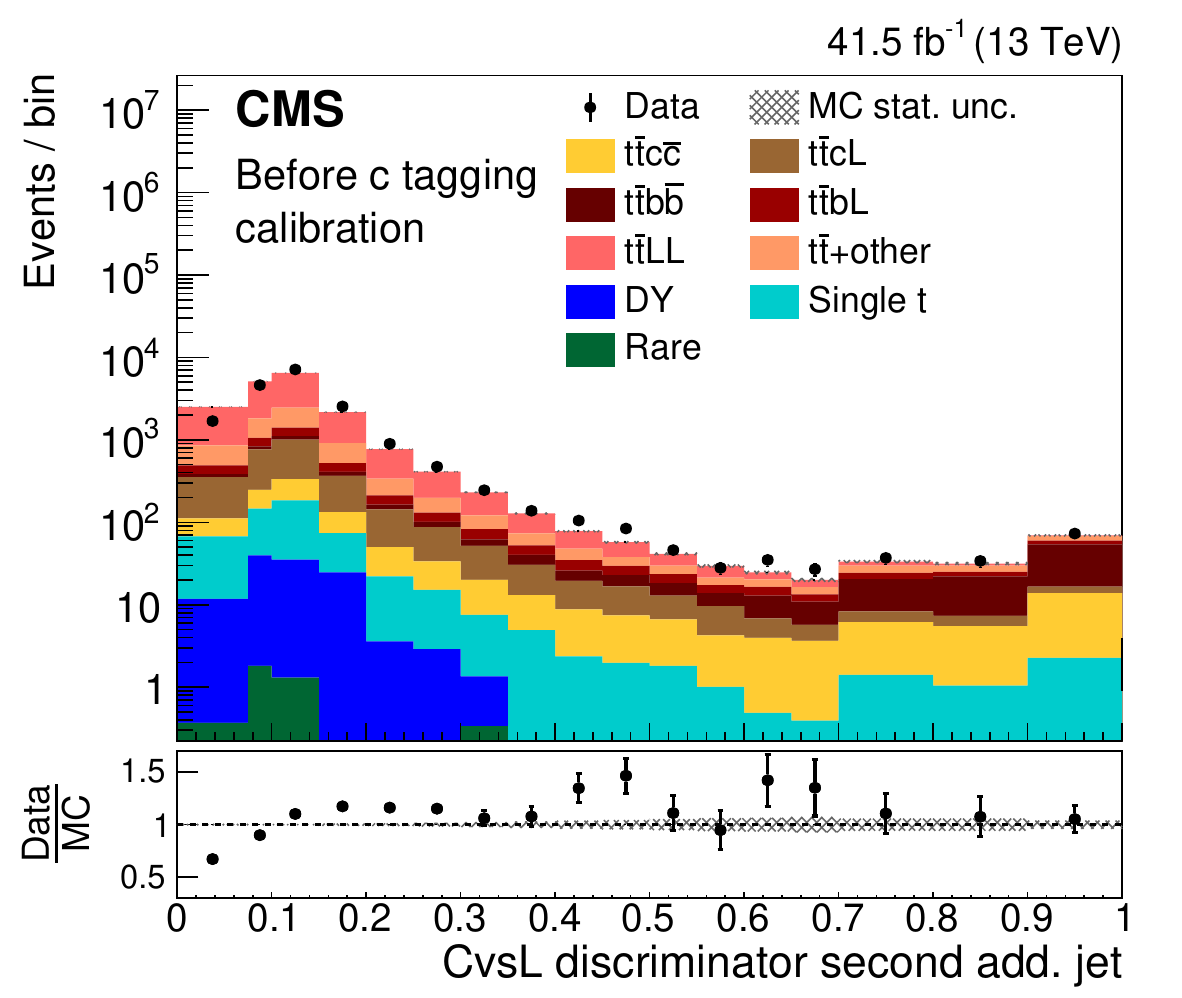}
  \includegraphics[width=.49\textwidth]{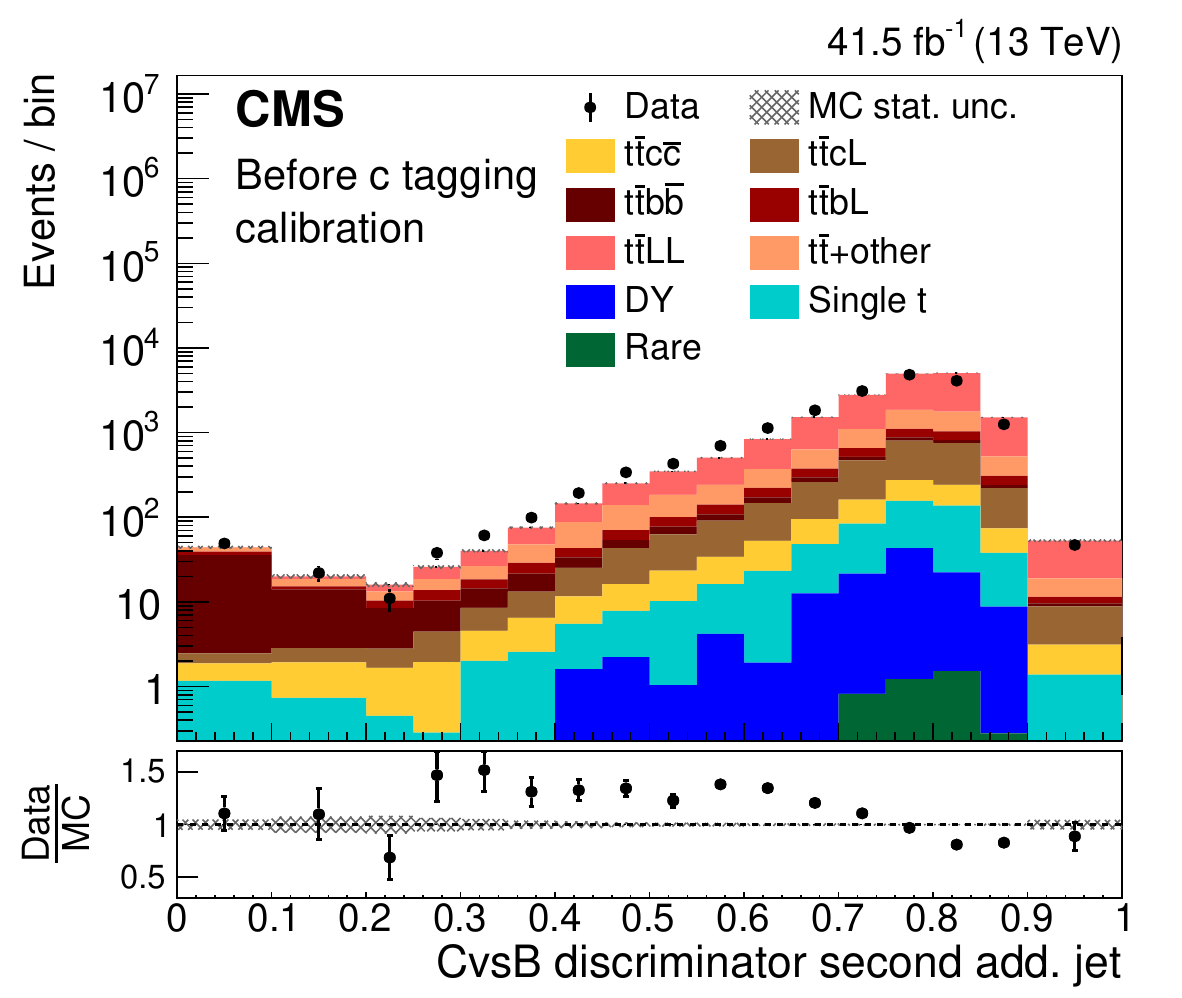}
  \caption{\label{fig:CTagBeforeCalibration}
  Comparison between data (points) and simulated predictions (histograms) for the CvsL (left column) and CvsB (right column) {\PQc} tagging discriminator distributions of the first (upper row) and second (lower row) additional jet before applying the {\PQc} tagging calibration. The lower panels show the ratio of the yields in data to those predicted in simulations. The vertical bars represent the statistical uncertainties in data, while the hatched bands show the statistical uncertainty in the simulated predictions.}
\end{figure*}

The {\PQc} tagging algorithms are trained on simulated events and are therefore prone to mismodelling effects in the input variables. The initial CvsL and CvsB {\PQc} tagging discriminator distributions of the first and second additional jet are shown in Fig.~\ref{fig:CTagBeforeCalibration}. Discrepancies between the data and the simulated predictions of up to 50\% are observed, demonstrating the need for the {\PQc} tagging calibration described in Section~\ref{sec:ctagCalib}.

\section{Charm tagging discriminators after scaling to the fitted cross sections.}\label{app:CtagDiscrAfterFitting}
\begin{figure*}[htb!]
\centering
\includegraphics[width=.49\textwidth]{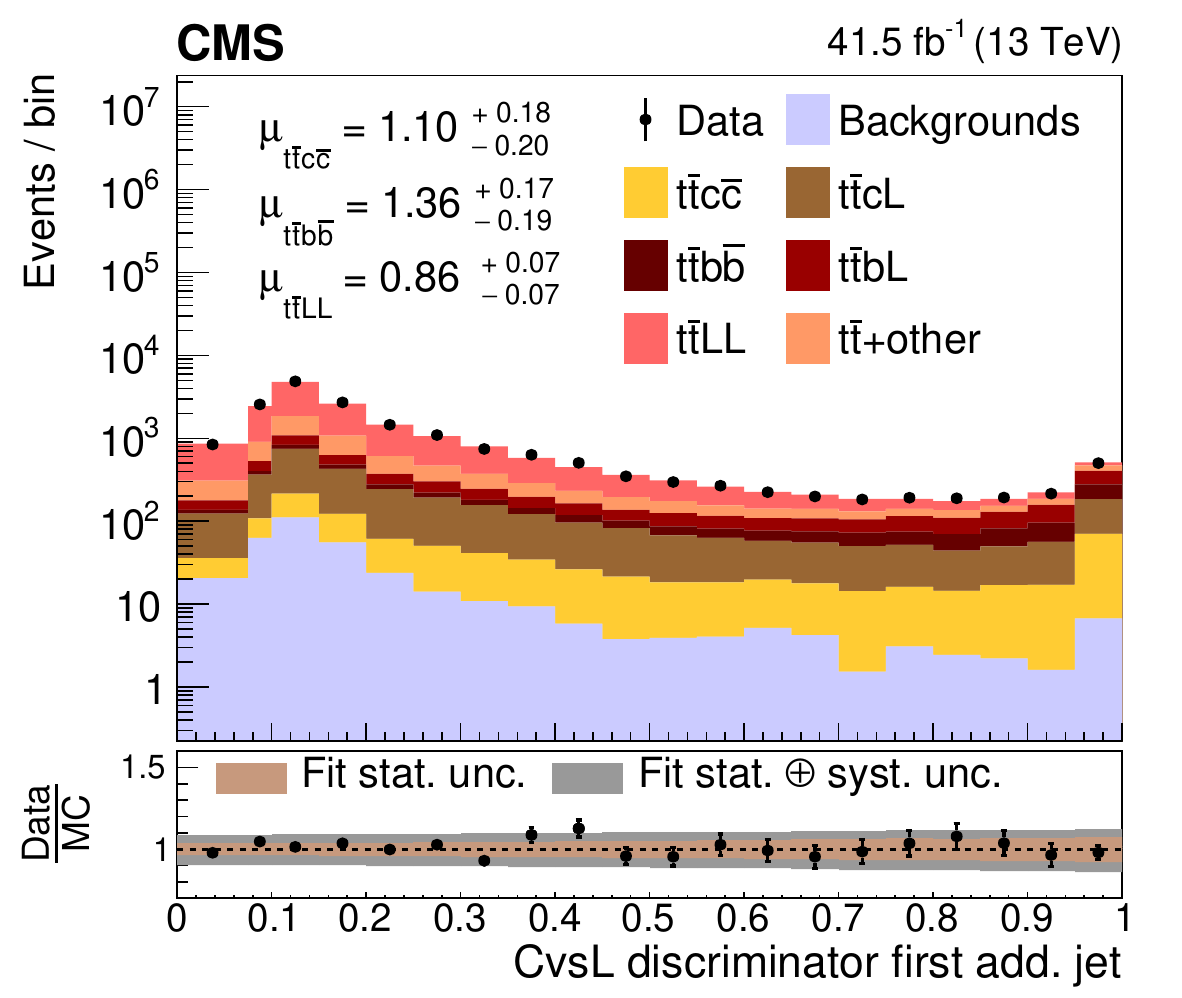}
\includegraphics[width=.49\textwidth]{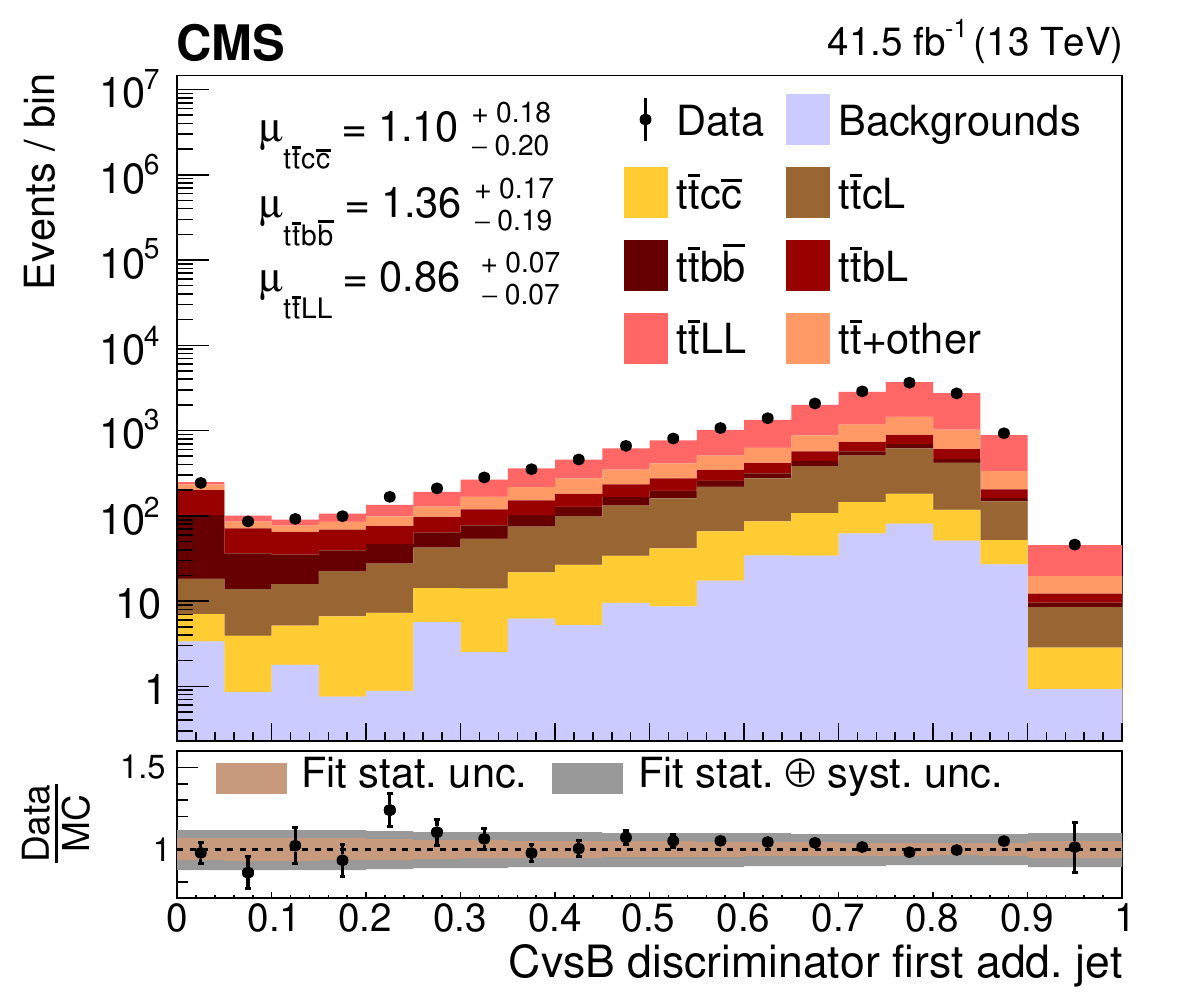}
\includegraphics[width=.49\textwidth]{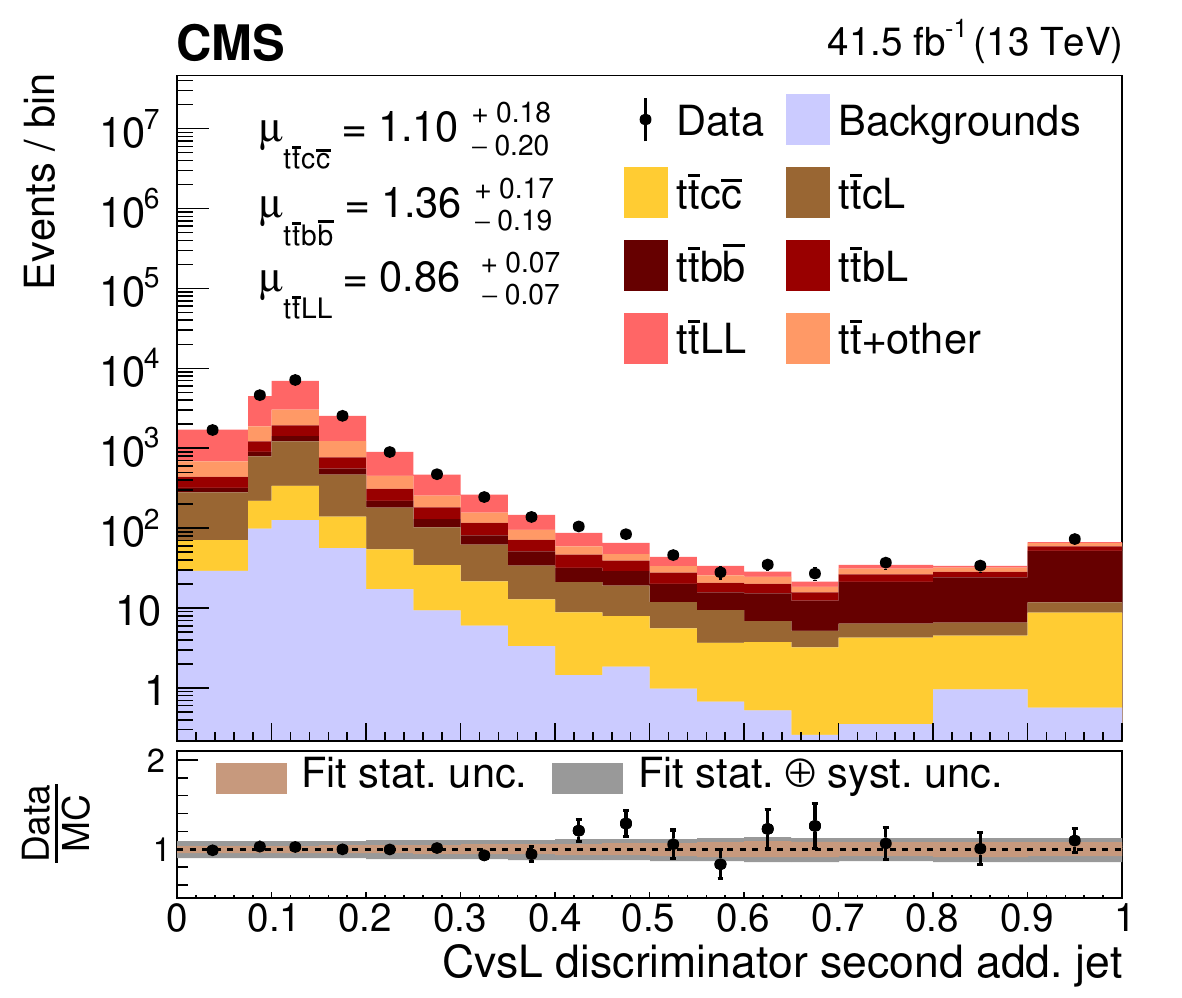}
\includegraphics[width=.49\textwidth]{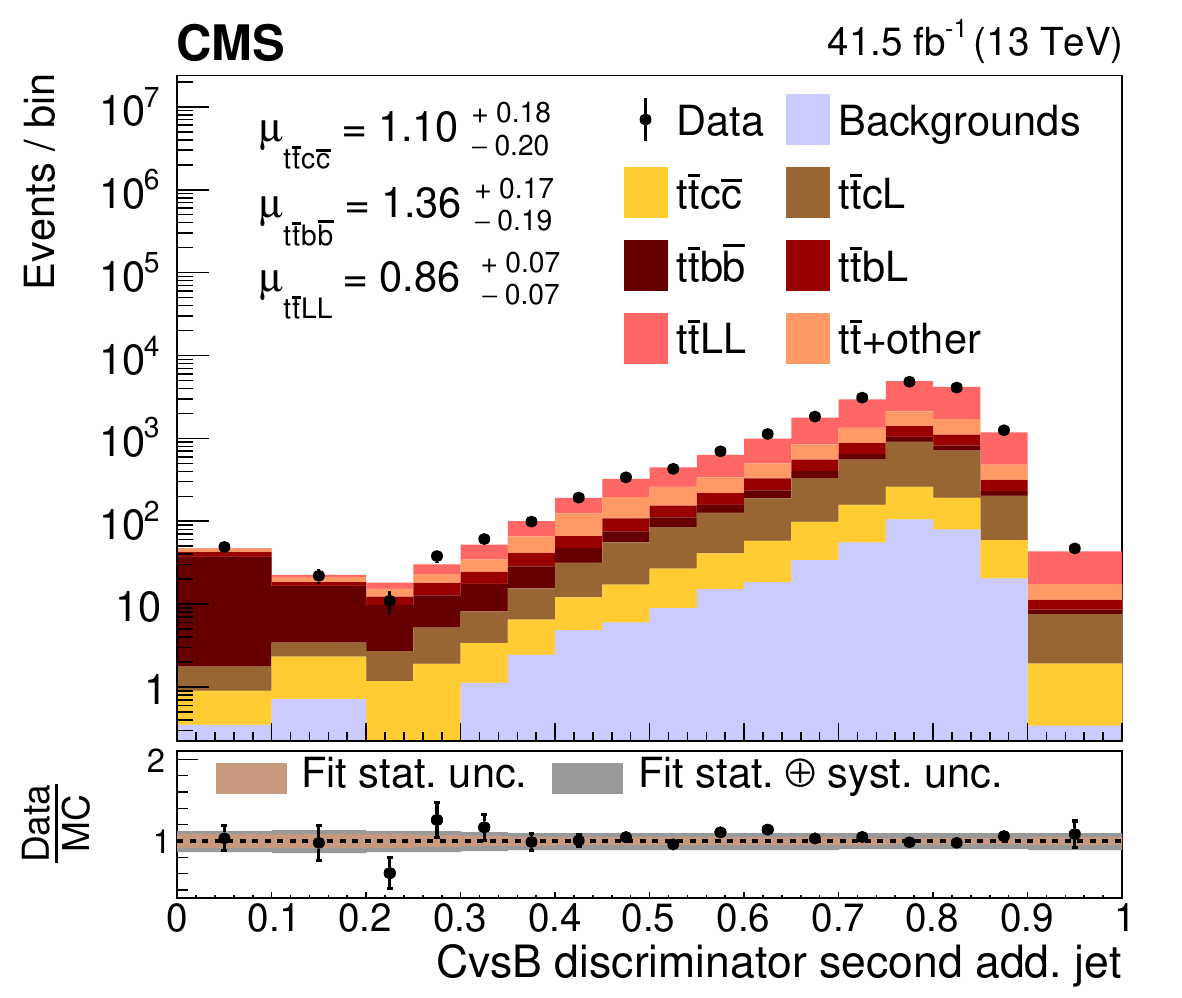}
\caption{\label{fig:CTagAfterFitting}
Comparison between data (points) and simulated predictions (histograms) for the CvsL (left column) and CvsB (right column) {\PQc} tagging discriminator distributions of the first (upper row) and second (lower row) additional jet, after normalizing the simulated templates according to the fitted cross sections. The lower panels show the ratio of the yields in data to those predicted in the simulations. The brown and grey uncertainty bands denote, respectively, the statistical and total uncertainties from the fit. The factors ($\mu$) by which the templates of the different processes (using the \POWHEG ME generator) are scaled, are also displayed, together with their combined statistical and systematic uncertainties.}
\end{figure*}

After scaling the simulated templates of the {\PQc} tagging discriminators to their fitted yields, the agreement between data and simulations is shown in Fig.~\ref{fig:CTagAfterFitting}.
This can be compared to the agreement before the fit was performed in Fig.~\ref{fig:CTagDilepton}, and indeed shows an improved agreement especially in those bins which have a relatively large contribution from $\ttbb$ and $\ttcc$ events.

\cleardoublepage \section{The CMS Collaboration \label{app:collab}}\begin{sloppypar}\hyphenpenalty=5000\widowpenalty=500\clubpenalty=5000\vskip\cmsinstskip
\textbf{Yerevan Physics Institute, Yerevan, Armenia}\\*[0pt]
A.M.~Sirunyan$^{\textrm{\dag}}$, A.~Tumasyan
\vskip\cmsinstskip
\textbf{Institut f\"{u}r Hochenergiephysik, Wien, Austria}\\*[0pt]
W.~Adam, T.~Bergauer, M.~Dragicevic, A.~Escalante~Del~Valle, R.~Fr\"{u}hwirth\cmsAuthorMark{1}, M.~Jeitler\cmsAuthorMark{1}, N.~Krammer, L.~Lechner, D.~Liko, I.~Mikulec, F.M.~Pitters, J.~Schieck\cmsAuthorMark{1}, R.~Sch\"{o}fbeck, M.~Spanring, S.~Templ, W.~Waltenberger, C.-E.~Wulz\cmsAuthorMark{1}, M.~Zarucki
% ...
\vskip\cmsinstskip
\dag: Deceased\\
1:  Also at Vienna University of Technology, Vienna, Austria\\
% ...
\end{sloppypar}
\end{document}